\documentclass{sig-alternate-2013}

\usepackage{balance}  
\usepackage{graphics} 
\usepackage{graphicx}
\usepackage{caption}
\usepackage{times}    
\usepackage{url}      

\usepackage{subfigure}
\usepackage{epstopdf}
\usepackage{epsfig}
\usepackage{amsmath}

\newcommand{\specialcell}[2][c]{
  \begin{tabular}[#1]{@{}c@{}}#2\end{tabular}}

\makeatletter
\def\url@leostyle{%
  \@ifundefined{selectfont}{\def\UrlFont{\sf}}{\def\UrlFont{\small\bf\ttfamily}}}
\makeatother
\def\pprw{8.5in}
\def\pprh{11in}

\usepackage[pdftex]{hyperref}
\hypersetup{
pdftitle={The Social Name-Letter Effect on Online Social Networks},
pdfauthor={LaTeX},
pdfkeywords={name-letter effect, Online Social Networks, same-name effect},
bookmarksnumbered,
pdfstartview={FitH},
colorlinks,
citecolor=black,
filecolor=black,
linkcolor=black,
urlcolor=black,
breaklinks=true,
}

\permission{Copyright is held by the International World Wide Web Conference Committee (IW3C2). IW3C2 reserves the right to provide a hyperlink to the author's site if the Material is used in electronic media.}
\conferenceinfo{WWW 2015,}{May 18--22, 2015, Florence, Italy.} 
\copyrightetc{ACM \the\acmcopyr}
\crdata{978-1-4503-3469-3/15/05. \\
http://dx.doi.org/10.1145/2736277.2741130}

\begin{document}

\title{Evolution of Conversations in the Age of Email Overload}

\numberofauthors{5}
\author{
\alignauthor Farshad Kooti\\
       \affaddr{USC Information Sciences Institute, USA}\\
       \affaddr{kooti@usc.edu}
\alignauthor Luca Maria Aiello\\
       \affaddr{Yahoo Labs Barcelona, Spain}\\
       \affaddr{alucca@yahoo-inc.com}
\alignauthor Mihajlo Grbovic\\
       \affaddr{Yahoo Labs Sunnyvale, USA}\\
       \affaddr{mihajlo@yahoo-inc.com}
\and
\alignauthor Kristina Lerman\\
       \affaddr{USC Information Sciences Institute, USA}\\
       \affaddr{lerman@isi.edu}
\alignauthor Amin Mantrach\\
       \affaddr{Yahoo Labs Barcelona, Spain}\\
       \affaddr{amantrac@yahoo-inc.com}
}

\maketitle


\begin{abstract}
Email is a ubiquitous communications tool in the workplace and plays an important role in social interactions. Previous studies of email were largely based on surveys and limited to relatively small populations of email users within organizations. In this paper, we report results of a large-scale study of more than 2 million users exchanging 16 billion emails over several months. We quantitatively characterize the replying behavior in conversations within pairs of users. In particular, we study the time it takes the user to reply to a received message and the length of the reply sent. We consider a variety of factors that affect the reply time and length, such as the stage of the conversation, user demographics, and use of portable devices. In addition, we study how increasing load affects emailing behavior. We find that as users receive more email messages in a day, they reply to a smaller fraction of them, using shorter replies. However, their responsiveness remains intact, and they may even reply to emails faster. Finally, we predict the time to reply, length of reply, and whether the reply ends a conversation. We demonstrate considerable improvement over the baseline in all three prediction tasks, showing the significant role that the factors that we uncover play, in determining replying behavior. We rank these factors based on their predictive power. Our findings have important implications for understanding human behavior and designing better email management applications for tasks like ranking unread emails.
\end{abstract}


\keywords{Emailing behavior, information overload, prediction}

\newcommand{\remove}[1]{}
\newcommand{\note}[1]{{\color{red} #1}}

\section{Introduction}\label{section:intro}

\noindent Electronic mail --- email ---  remains an essential tool for social interactions and a popular platform for computer-mediated communication. It is used within organizations to exchange information and coordinate action, but also by ordinary people to converse with friends. Patterns of email interactions reveal circadian rhythms~\cite{Malmgren08pnas} and bursty dynamics of human activity~\cite{Barabasi05nature}, and the structure of evolving conversations~\cite{Eckmann04}. Understanding how these patterns shape email use is necessary for designing the next generation of interaction tools that will improve the efficiency of communication and coordination in social groups.

Early studies of email examined how people process~\cite{Dabbish05,Dabbish06}, organize~\cite{Whittaker96}, and respond~\cite{Neustadter05} to email messages. By surveying users within organizations, researchers uncovered common email triage strategies, with some users processing emails serially, while others read and reply to important messages first~\cite{Neustadter05}.

Those studies had relatively small sample sizes and were limited by their methodology to answering qualitative questions about email behavior. As a result, we do not have answers to a number of questions about how people use email. How many email conversations do people have? How long are these conversations and how do they end? When do people respond to a message in a conversation? Do they adapt their replies to the behavior of their conversation partner?

Furthermore, as the volume of email has increased steadily over the years, the concept of ``email overload'' has grown in prominence. Dabbish \& Kraut~\cite{Dabbish06} defined ``email overload'' as \textit{``email users' perceptions that their own use of email has gotten out of control because they receive and send more email than they can handle, find, or process effectively.''}. This motivates a whole set of questions that have been recently addressed in the setting of online social networks~\cite{goncalves11modeling,gomez14quantifying}, but never fully in the context of email. How does the volume of incoming email, information or email load, affect user behavior? How do people compensate for the increased load: do they take longer to reply, or do they send shorter replies that take less time to compose?

We address these questions with a largest study to date of email conversations (16B emails). We focus our analyses on the replying behavior within \textit{dyadic interactions}, i.e., conversations between pairs of users. Specifically, we measure the time a user takes to reply to a message, the length of the reply, as well as the fraction of messages a user replies to. First, we empirically characterize replying behavior in email conversations in a large population of users, and also how these behaviors vary by gender and age. Although we find no significant variation due to gender, we find that younger email users reply faster and write shorter replies than older users.

Next, we study how email load, measured by the number of received email messages, affects replying behavior. We find that while users attempt to adapt to the rising information load by replying to more emails, they do not adequately compensate. As email load increases, they reply faster but to a decreasing fraction of incoming emails. These findings suggest that email overload is a problem, with users generally unable to keep up with rising load.

We also study how replying behavior evolves over the course of a conversation. We find that users initially synchronize their behaviors, with reply times and length of replies becoming more similar in the first half of the conversation. However, they become desynchronized in the second half of the conversation. In contrast, users continue to coordinate their linguistic styles, as messages exchanged over the course of the entire conversation become more similar in content and style.

Finally, we develop a model to predict users' replying behavior in a conversation, namely, how long it will take the user to reply to a message, the length of the reply, and if the reply will end the thread. The features that we use for prediction include history of  communication between the users, user demographics, email load, as well as the day of the week and time the message was received. Our predictive model considerably outperforms the baseline in all three cases. We obtain accuracy of 58.8\% for reply time, which is a 67.1\% relative improvement over the baseline, and 71.8\% accuracy for length of replies (113.7\% relative improvement over the baseline). We also predict the last email of the thread with accuracy of 65.9\%,  a 30.2\% relative improvement over the baseline. Ability to accurately predict what messages a user will reply to can be used by email clients to rank emails in the users' inbox by their replying priority, thus helping ease the burden of information overload.

The key contributions of our work are:
\begin{enumerate}
\item We empirically characterize email \textit{replying behavior} of users, focusing on reply time, length of the reply, and the correlation between them. We quantify how different factors, including the day and time the message was received, the device used, the number of attachments in the email, and user demographics affect replying.
\item We show that \textit{email overload} is evident in email usage and has adverse effects, resulting in users replying to a smaller fraction of received emails. Users tend to send shorter replies, but with shorter delays when receiving many emails. We find that different age groups cope with overload differently: younger users shorten their replies, while older users reply to smaller fraction of received messages.
\item We find evidence of \textit{synchronization} in dyadic interactions within a thread: users become more similar in terms of reply time and length of replies until the middle of a thread, and start acting more independently after that.
\item We can predict reply time and length, and the last email in a thread with a much higher accuracy than the baseline. This has important implications for designing future email clients.
\end{enumerate}

\section{Dataset}\label{section:dataset}

\noindent Yahoo Mail is one of the largest email providers in the world, with over 300M users (according to ComScore\footnote{\small{http://www.comscore.com}}), who generate an extremely high volume of messages. Obviously, not all the email addresses are associated with real people and are used instead by organizations or possibly bots to generate emails for commercial promotions and spam~\cite{jagatic07social,zhuang08characterizing,grbovic2014many}. To meet the goal of studying social interactions, it is necessary to subsample the data to a set of interactions that are likely occurring between real people. For this reason, we apply a conservative filtering strategy and focus our study on user pairs (or \textit{dyads}) that exhibit \textit{reciprocal interaction} (i.e., bi-directionality of emails sent) and exchange some minimum number of messages. This ensures that \textit{i)} all the email addresses of the dyadic endpoints are likely associated with human users, and that \textit{ii)} the emails sent between them are not automatically generated.

Accordingly, we selected a random subsample $\mathcal{E}$ of 1.3M dyads of Yahoo Mail users worldwide who have significant interactions with each other, sending at least five replies in each direction in the time span of undisclosed number of months. For privacy and policy reasons, the dataset includes messages belonging exclusively to users who voluntarily opted-in for such studies. Consequently, considering a pair of users who exchanged more than 5 emails, both users need to be opt-in users to be included in the study. These pairs comprise a set $\mathcal{N}$ containing 2M unique users exchanging 187M emails overall. We refer to the full sequence of emails flowing within the dyad as a \textit{conversation}. Next, we gathered \textit{all} incoming and outgoing emails of users in $\mathcal{N}$ over the same time period, a total of 16B emails. Note that these emails included only emails from commercial domains and emails to and from other opt-in Yahoo users. Due to Yahoo policy, the study did not include personal email messages between Yahoo users and other email clients. In addition, we excluded notifications from social network sites, such as Facebook or Twitter, which represent a considerable portion of commercial, automatically-generated emails.

We conduct our study at two different levels: at the dyadic level, we consider only the emails flowing between dyad endpoints; at the global level, we consider the entire dataset. In the latter case, commercial and spam messages will be likely be part of the incoming emails directed to users in $\mathcal{N}$. As one of the goals of this work is to study information overload in all its facets, we don't apply any \emph{a priori} filter. Then, depending on the target of each part of the analysis, we apply ad-hoc filters to fit our specific goals.


Each email included the \textit{sender ID}, \textit{receiver ID}, \textit{time sent}, \textit{subject}, \textit{body} of the email, and number of \textit{attachments}. All data were anonymized to preserve user privacy and worked at the level of anonymized user ids and email ids. Given the data sensitivity, even with opt-in users, email bodies were not editorially inspected by humans. To extract statistics from email bodies (length, number of articles, email vectors, email ids in a thread, etc.), we made a MapReduce call with a specified function to a protected cluster.


\begin{figure}[t!]
\begin{center}
\includegraphics[width=0.99\columnwidth]{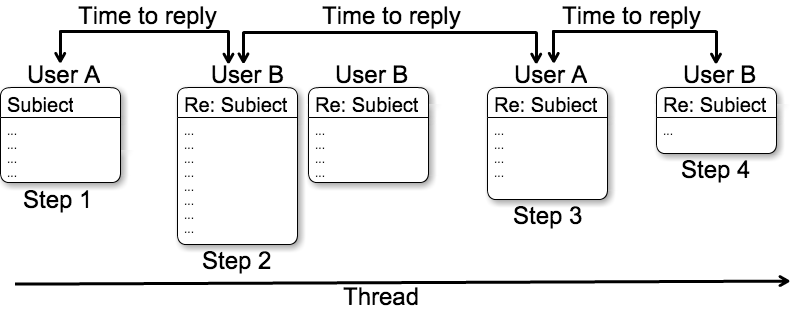}
\end{center}
\vspace*{-2mm}
\caption{Illustration of an email thread.}
\label{fig:diagram}
\vspace*{-2mm}
\end{figure}

To make sense of the conversation flow, we need to break down the dyadic conversations into \textit{threads}. A thread is an ordered sequence of one or more emails, including the initial email, and the list of succeeding replies to it (if any). As our dataset does not keep track of the thread structure, we need to reconstruct it. We compose the threads using the subject line and the time each email was sent. Yahoo Mail automatically adds the token \textit{``Re: ''} to the subject line of all replies. For each pair of the users, we group all the exchanged emails that have the same subject line. If all the emails in the group start with ``Re:'', we also add the email that has the exact subject without the ``Re:''; this email would be the first email of the thread. Then, in each group we order the emails based on the sent time, and the time to replies could be calculated as the delay between the email from one user and the reply of the other user. In case of consecutive replies in one thread from one of the users, we just consider the first one (Figure~\ref{fig:diagram}).

Like most other email service providers, Yahoo Mail quotes the body of the original message at the end of the reply, unless manually removed. To get only the text of the last email sent, we search for standard string templates that occur before the quoted message, (e.g., ``On Thursday May 1, 2014 a@yahoo.com wrote'') and exclude any text from that point. We also looked for common mobile device signatures, such as "Sent from my iPhone" and exclude the text from the signature onwards.

\vspace{5pt} \noindent {\bf Limitations.} Our data comes with a few limitations. First, in case of an in-line reply, we would not be able to detect the quoted message, and it would be considered as a new message. The quoted message could be caught by comparison of reply and the email being replied to, but the string matching would be computationally expensive for the scale of our dataset. Second, we are not able to handle the emails that are sent to a group of users, and we consider all the emails as being dyadic. Moreover, for constructing the threads, we use the exact subject line and if there are two different threads between two users with the same subject, we will consider both of them as one thread. We believe these cases are special cases and a small fraction of emails, so our results would not be effected by these limitations.

\vspace{5pt} \noindent {\bf Spam.} One main concern with studying emails is the spam. To minimize the effect of spam, we conduct most of our analyses on dyadic email exchanges that occur between two users who have exchanged at least five emails with each other. We believe that the fact that two users have exchanged more than five emails means that neither of them is a spammer. In the analysis of the information load on users, we have to consider all the emails those users received, and many of them could be spam and should not be considered. To deal with this problem, we conduct the analysis once for all the received emails and once for emails received from contacts, i.e., others with whom the user had exchanged at least one email. Again, filtering users who have not received any replies from a user is a very conservative approach to eliminate spammers.

\begin{figure}[t!]
\begin{center}
\includegraphics[width=0.7\columnwidth]{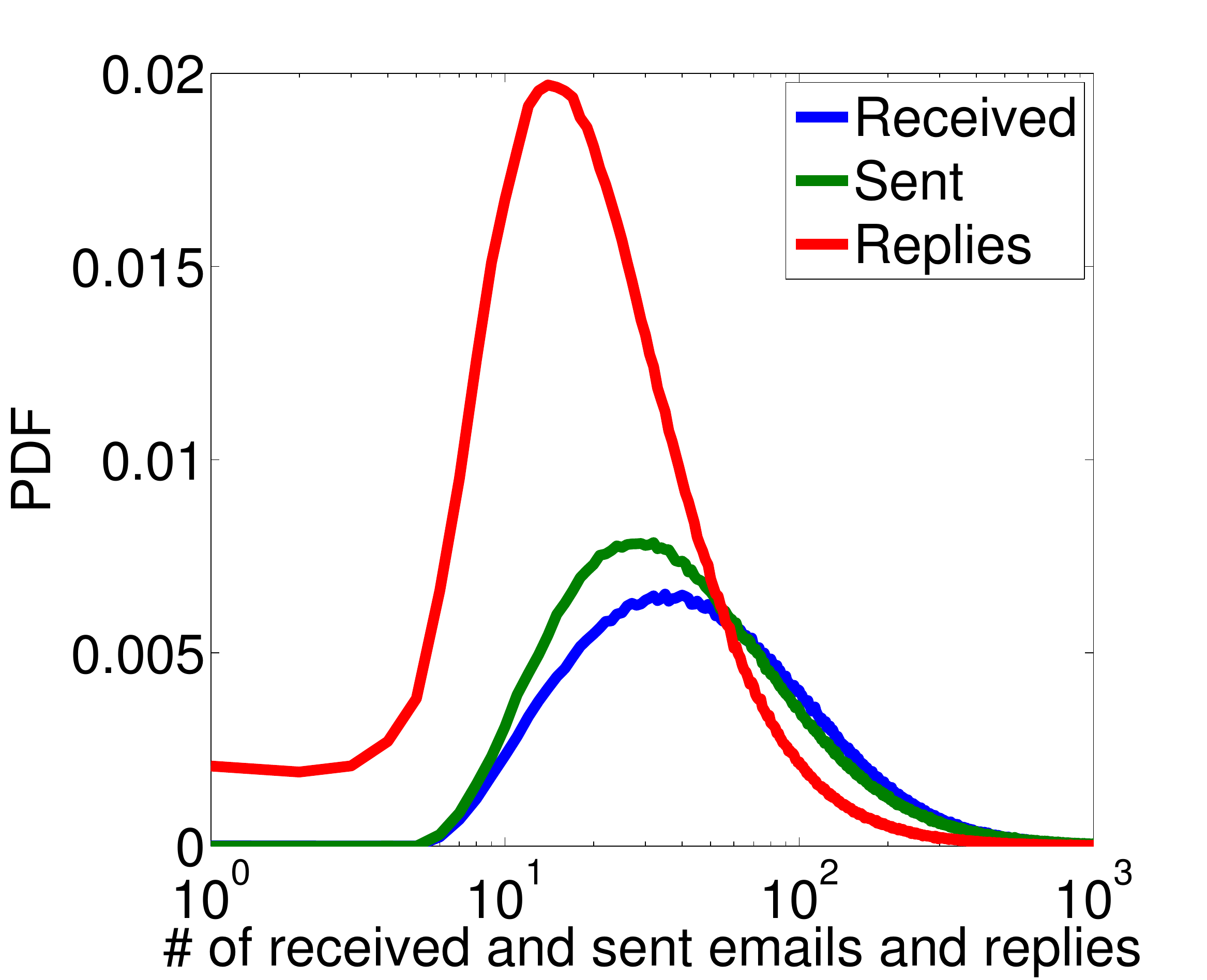}
\end{center}
\caption{Distribution of number of received, sent, and replied emails.}
\label{fig:rec_sent_reply}
\vspace*{-4mm}
\end{figure}

\begin{figure}[t!]
\begin{tabular}{@{}c@{}c@{}}
\subfigure[Number of emails in a thread]{
   \includegraphics[width=0.5\columnwidth]{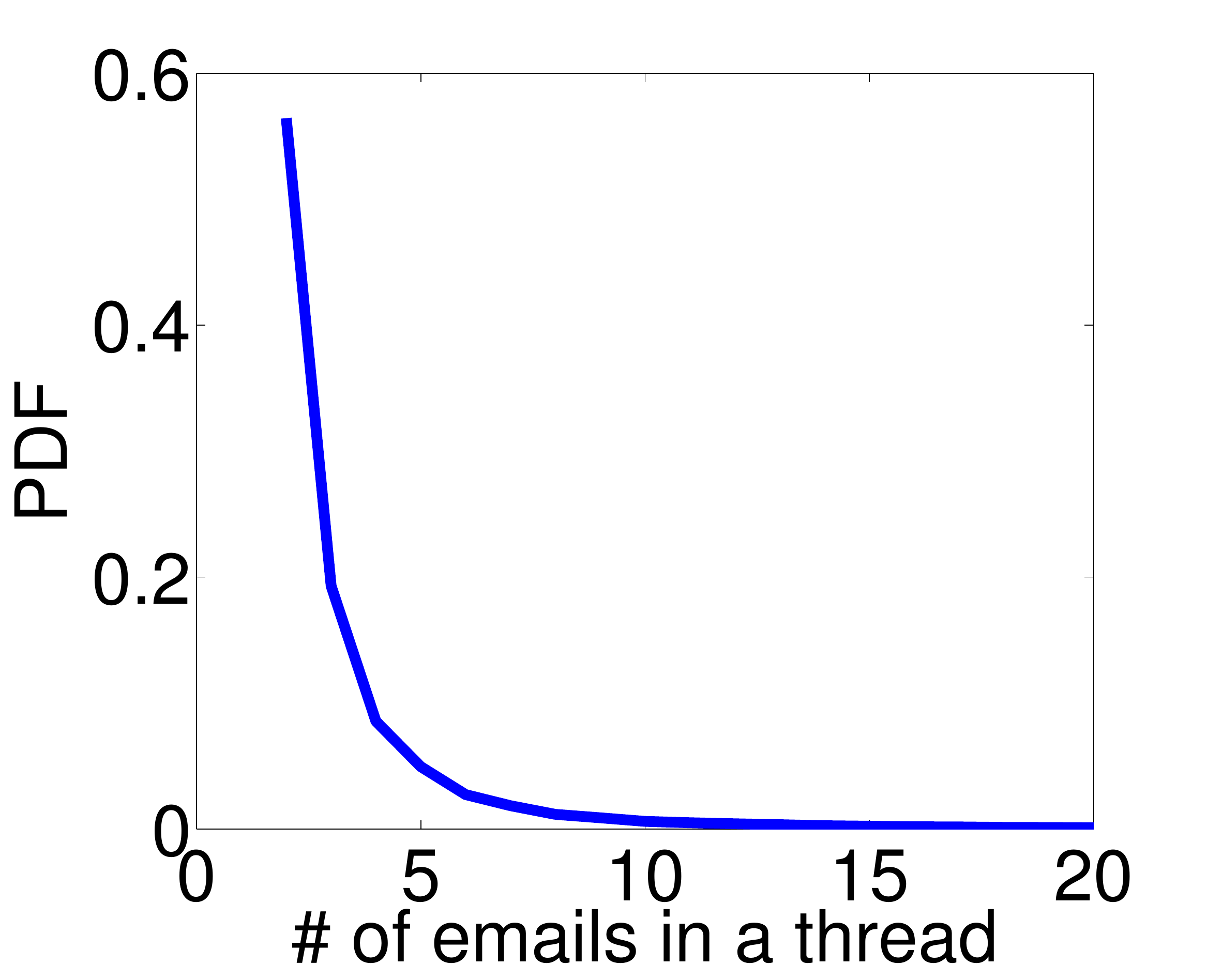}
      \label{fig:thread_dist_pdf}
   }
  &
   \subfigure[Number of threads]{
   \includegraphics[width=0.5\columnwidth]{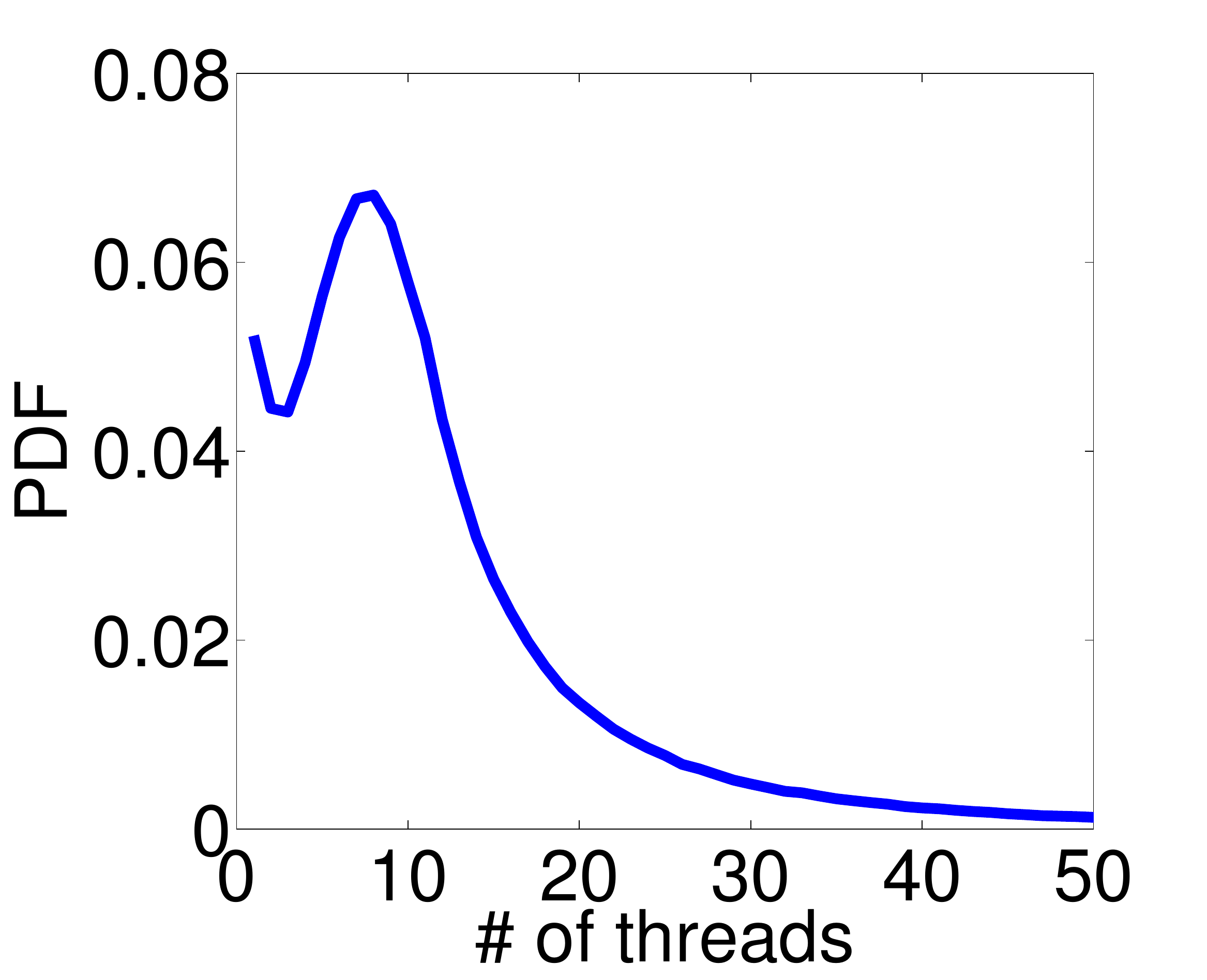}
   \label{fig:thread_in_conv_pdf}
   }
  \end{tabular}
  \vspace*{-1mm}
   \caption{Statistics of conversations. (a) Number of emails in a thread (mean 3.76 emails, median 2 emails) and (b) the number of threads in conversations (mean 13.94 emails, median 9 emails).}
  \label{fig:conversations}
  \vspace*{-2mm}
\end{figure}

\begin{figure}[t!]
\begin{center}
\includegraphics[width=0.70\columnwidth]{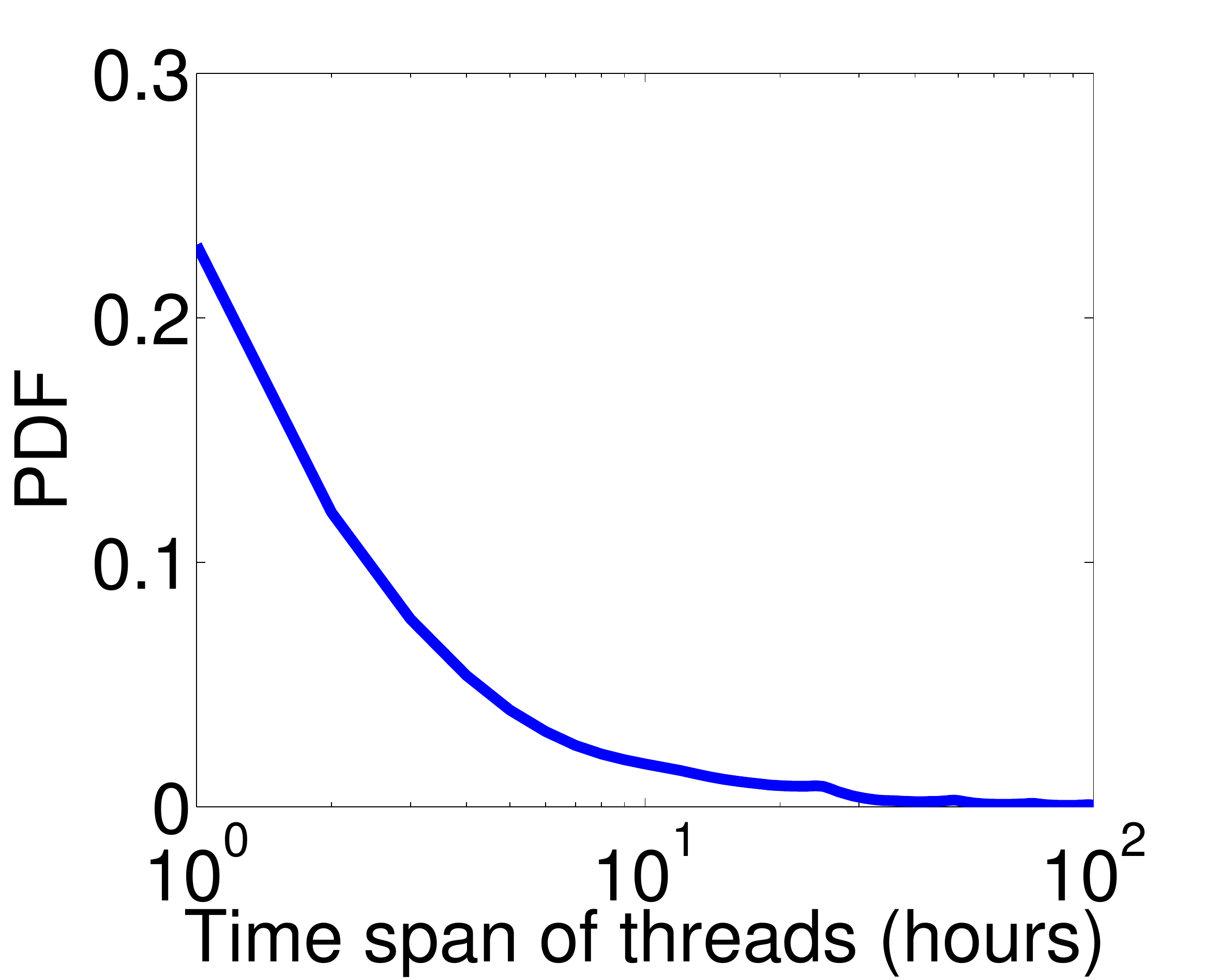}
\end{center}
\vspace*{-2mm}
\caption{Distribution of time span of threads (mean 53.2 hours, median 3.5 hours).}
\label{fig:thread_time}
\vspace*{-3mm}
\end{figure}

Figure~\ref{fig:rec_sent_reply} shows the distribution of the number of received, sent, and replied email messages for the 1.3M users for whom we have incoming and outgoing emails outside $\mathcal{E}$. Figure~\ref{fig:thread_dist_pdf} shows the distribution of the number of emails in a thread across all conversations. This distribution has the expected heavy-tailed shape: most threads are very short, e.g., more than 30\% of threads have only one step, which means, it's an email and a single reply to it. We also look at the number of the threads in a conversation (i.e., all the emails exchanged) during the considered period (Figure~\ref{fig:thread_in_conv_pdf}). There is a small drop at the beginning, and then an increase up to 10 threads, after that users are less likely to have more threads. Also to better understand the time scale of communications, we measure the longevity of threads for all users (Figure~\ref{fig:thread_time}). Many threads last for only a few hours; half of the threads last 3.5 hours or less. But, there is a considerable fraction of threads that last longer a day (22\%).

\section{Replying Behavior}\label{section:characterization}

\noindent We characterize the replying behavior of email users involved in dyadic interactions, focusing on reply time and the length of replies.

\subsection{Reply time and length}

\noindent Reply time is the period between the time the sender sends a message (e.g., ``user A'' in Figure~\ref{fig:diagram}) and the time the receiver (``user  B'') replies to it. When receiver replies after multiple consecutive emails are sent by the sender within the same thread, reply time is calculated from the time of the first message to the time the receiver replied. We experimented with different definitions of reply time, e.g., from the time of the last messages in a series of emails in a thread, but this did not significantly change the key properties of replying behavior, resulting only in slightly faster replies.

\begin{figure}[t!]
\begin{tabular}{@{}c@{}c@{}}
\subfigure[PDF]{
   \includegraphics[width=0.5\columnwidth]{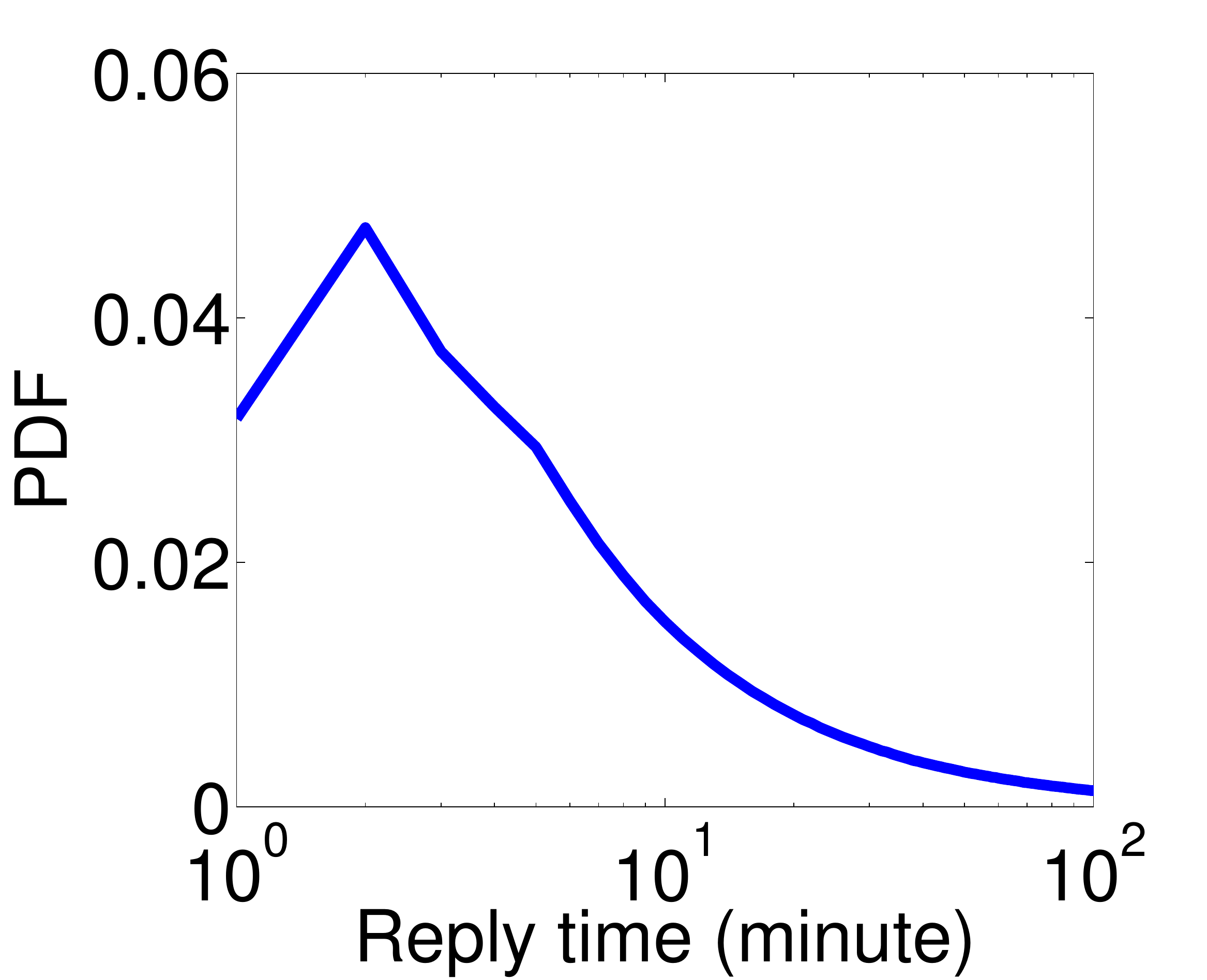}
      \label{fig:time_all_pdf}
   }
  &
   \subfigure[CDF]{
   \includegraphics[width=0.5\columnwidth]{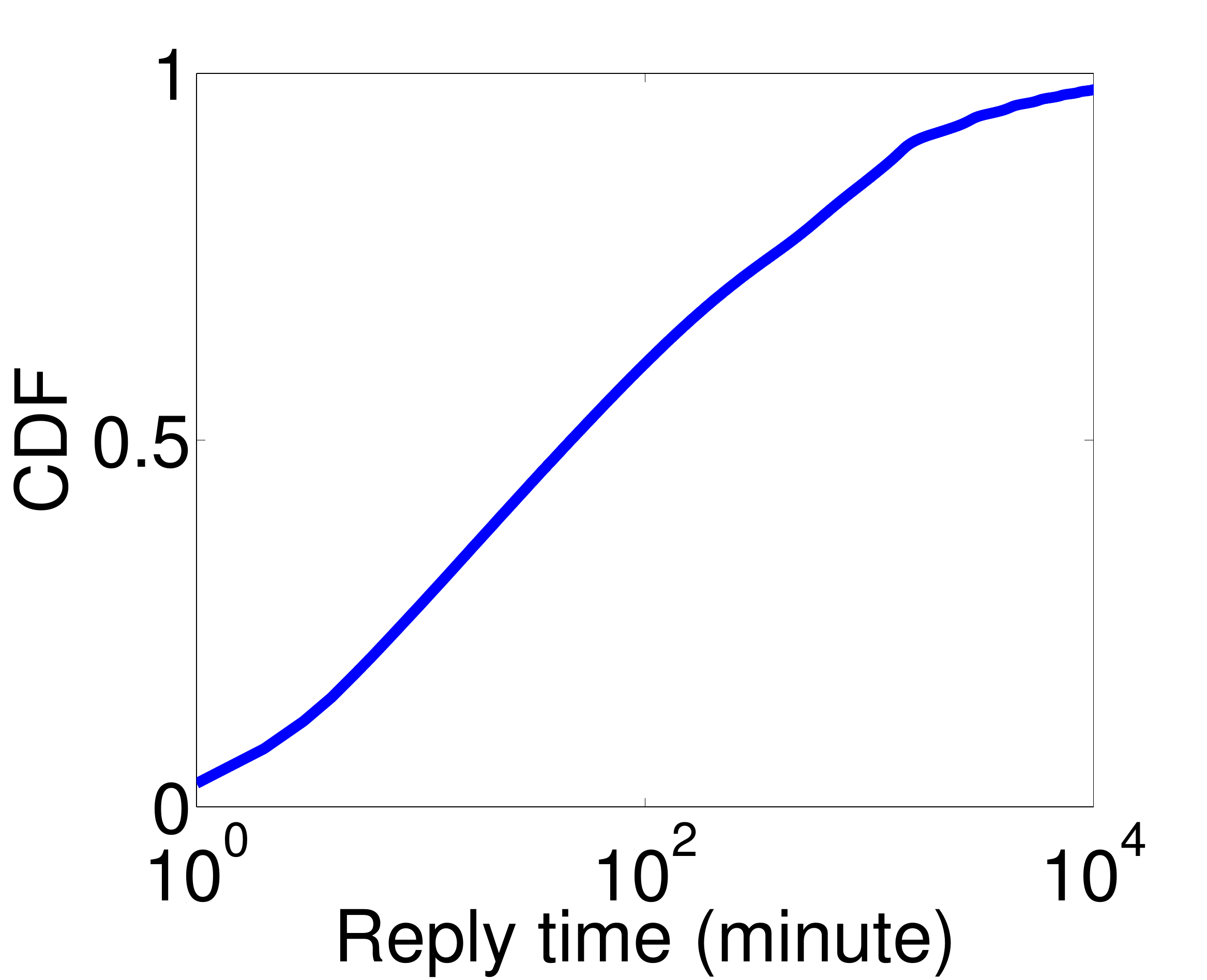}
   \label{fig:time_all_pdf}
   }
  \end{tabular}
   \caption{Distribution of reply times. In general, replies are very fast (mean:1157 minutes, median: 47 minutes, standard deviation: 19730).}
  \label{fig:time_to_reply}
   \vspace*{-3mm}
\end{figure}

Figure~\ref{fig:time_to_reply} shows the probability distribution function (PDF) and the cumulative distribution function (CDF)  of reply times. Most of the replies are very fast: more than 90\% happen within a day of receiving the message, and the most likely reply time is just two minutes. Also, half of the replies are within 47 minutes of receiving the message. Interestingly, this distribution is very similar to the time it takes users to retweet a message on Twitter~\cite{Hodas12socialcom}.

\begin{figure}[t!]
\begin{tabular}{@{}c@{}c@{}}
\subfigure[PDF]{
   \includegraphics[width=0.5\columnwidth]{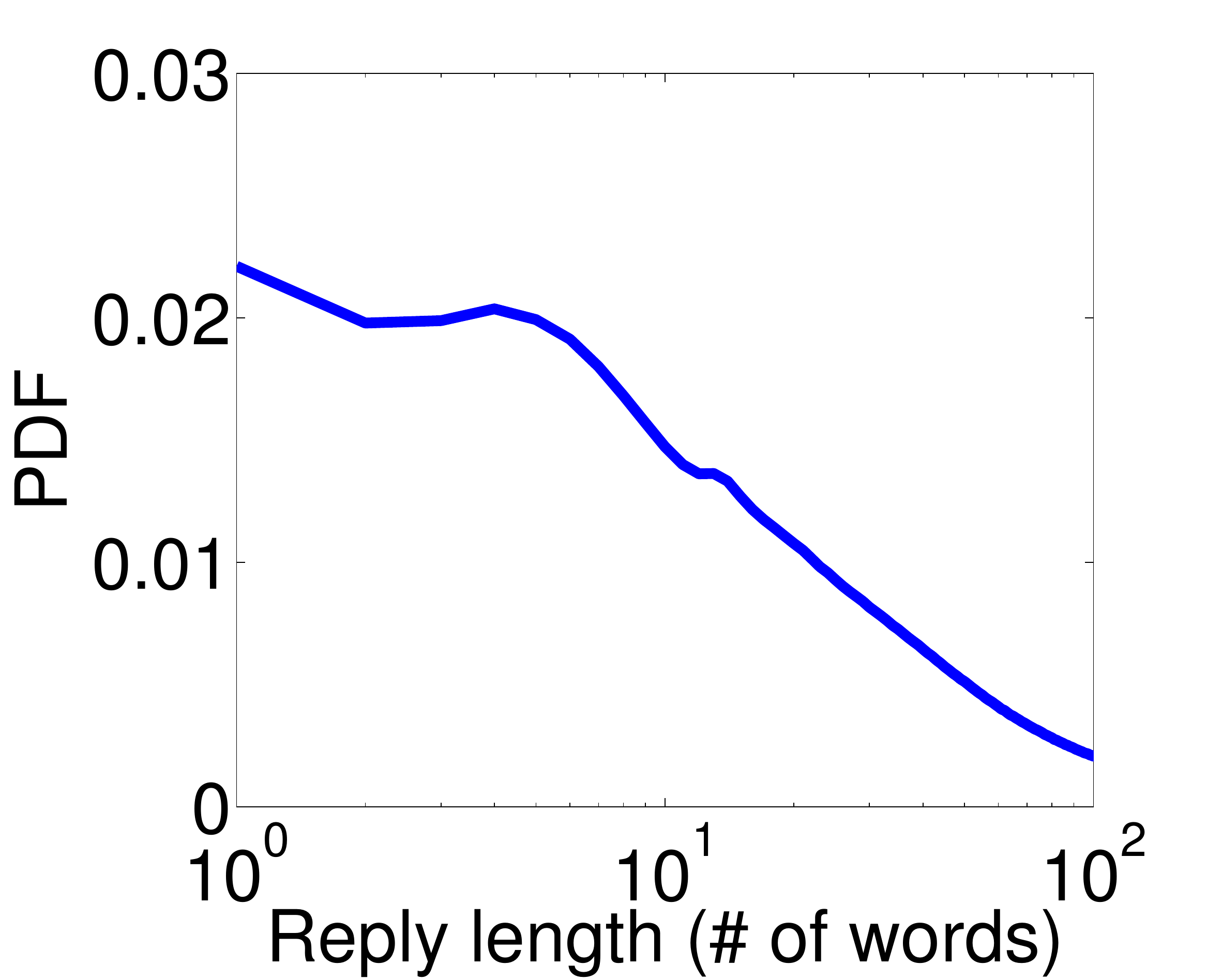}
      \label{fig:length_all_pdf}
   }
  &
   \subfigure[CDF]{
   \includegraphics[width=0.5\columnwidth]{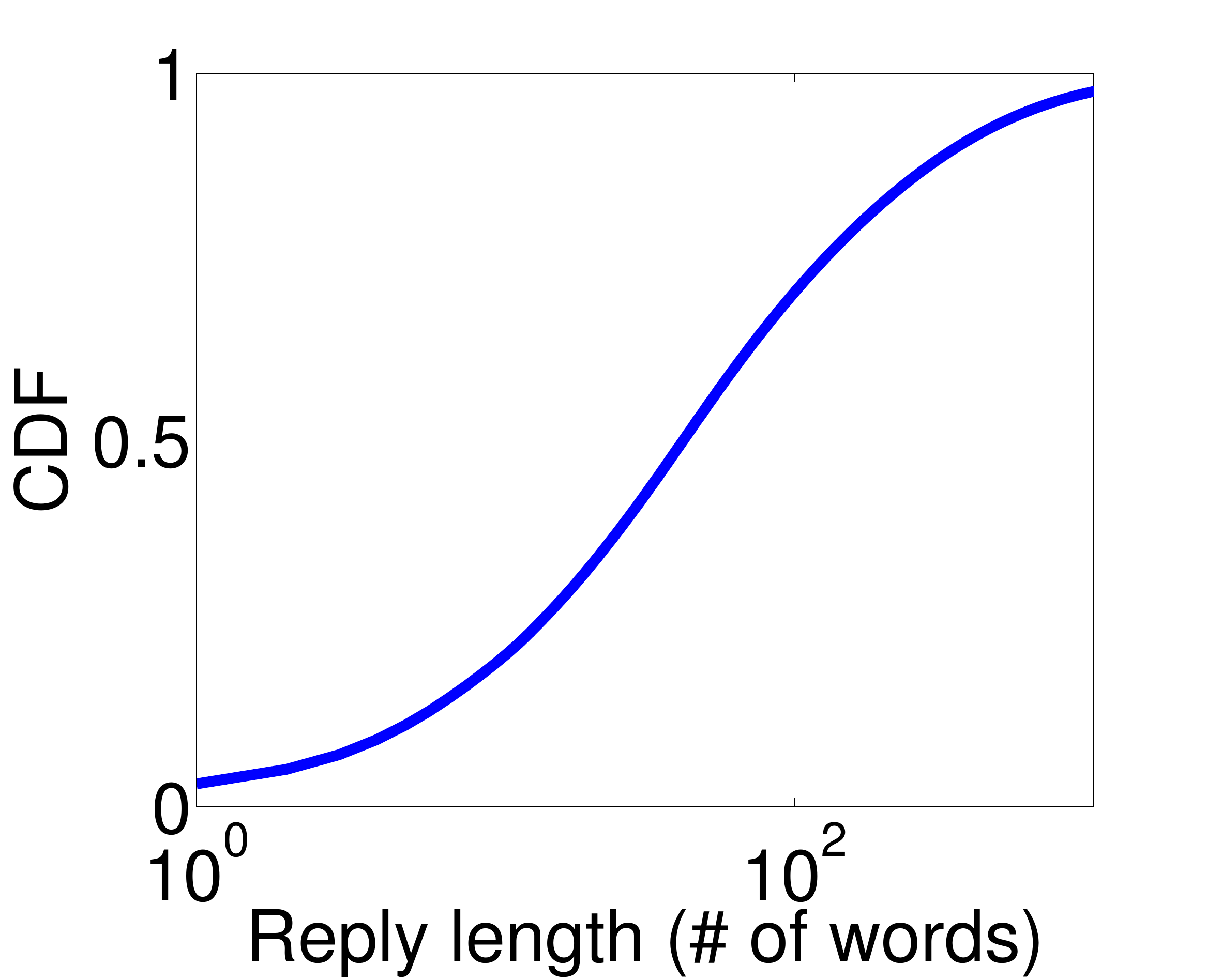}
   \label{fig:length_all_pdf}
   }
  \end{tabular}
   \caption{Distribution of reply lengths (mean 153 words, median 43 words, standard deviation 419).}
  \label{fig:length_of_reply}
   \vspace*{-3mm}
\end{figure}

The length of a reply is the length of the message receiver sends back to the sender, excluding the body of the original message, which may have been quoted in the email. Figure~\ref{fig:length_of_reply} shows the PDF and CDF of the length of replies. Considerable fraction of replies are very short: the most likely reply length is only five words, with half of the replies being shorter than 43 words. However, there is a non-negligible fraction of long emails; 30\% of emails are longer than 100 words.

\begin{figure}[t!]
\begin{tabular}{@{}c@{}c@{}}
\subfigure[Time to reply]{
   \includegraphics[width=0.5\columnwidth]{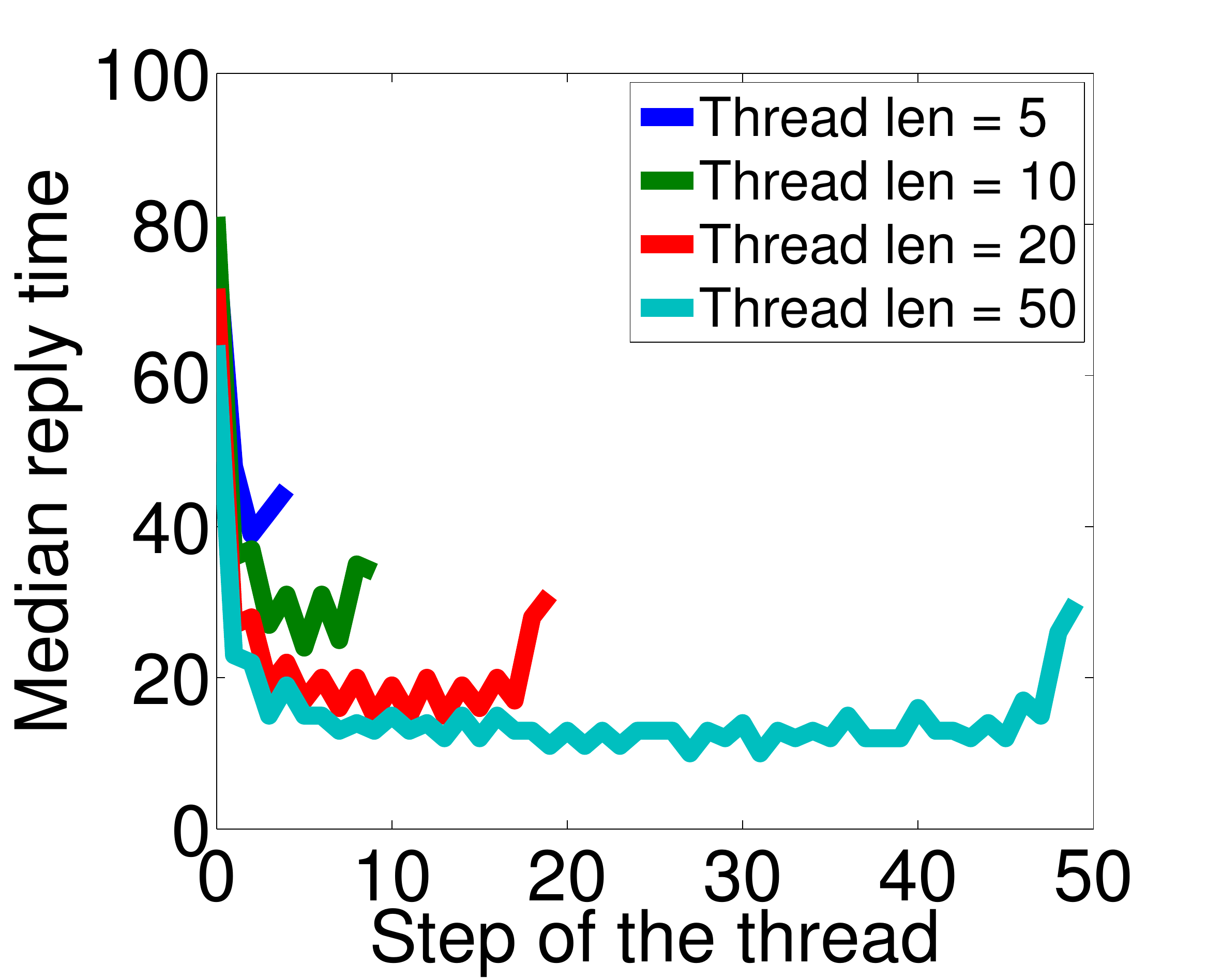}
      \label{fig:time_step}
   }
  &
   \subfigure[Length of reply]{
   \includegraphics[width=0.5\columnwidth]{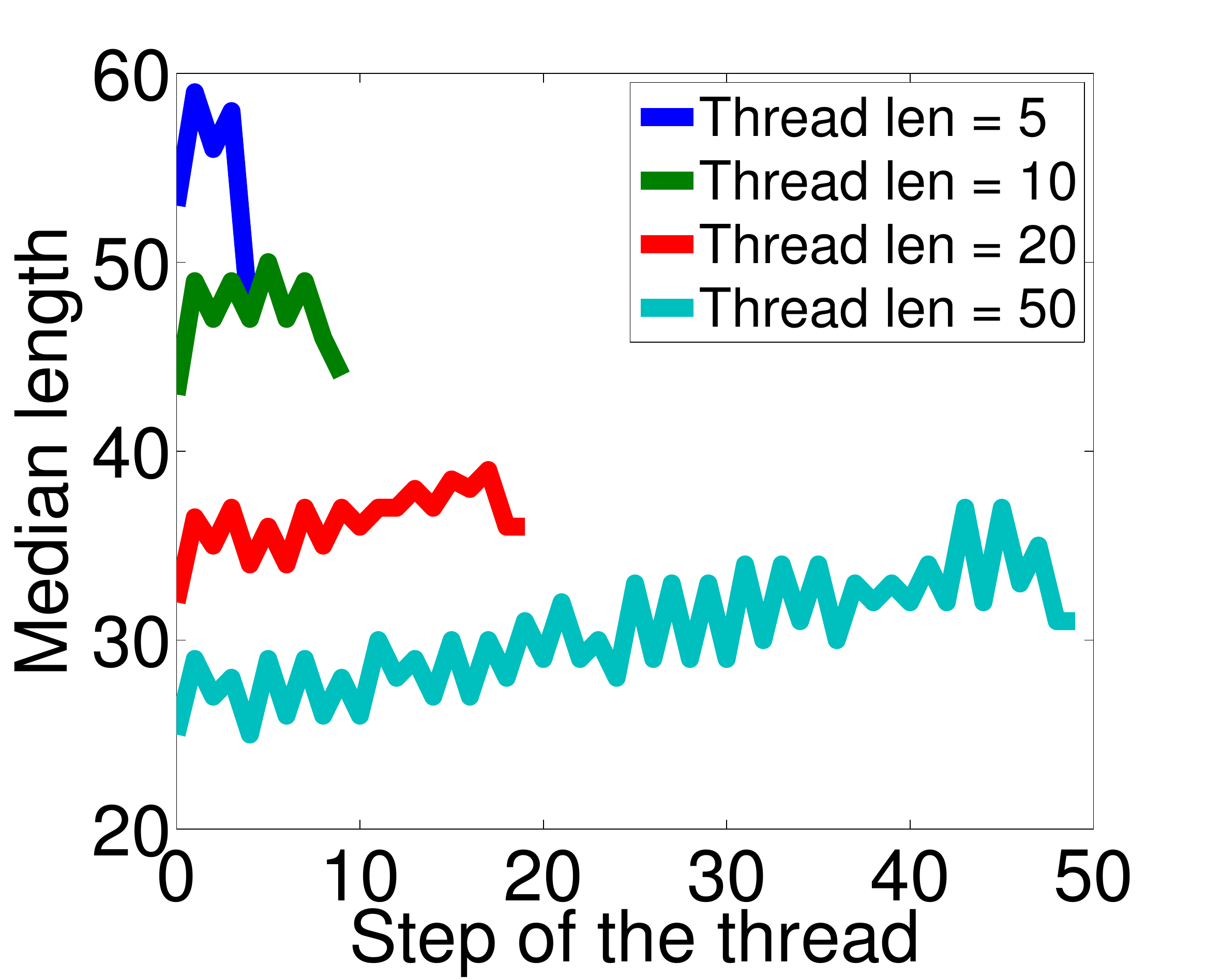}
   \label{fig:length_step}
   }
  \end{tabular}
   \caption{(a) Median reply time for different steps of threads for a given thread length. Replies become faster, except for the very last reply that is much slower. (b) Median length of reply for different steps of threads for a given thread length. Calculated on dyadic conversations.}
  \label{fig:time_length_step}
   \vspace*{-2mm}
\end{figure}

\subsection{Evolution of Conversations}
\begin{figure}[t!]
\begin{tabular}{@{}c@{}c@{}}
\subfigure[Reply time]{
   \includegraphics[width=0.5\columnwidth]{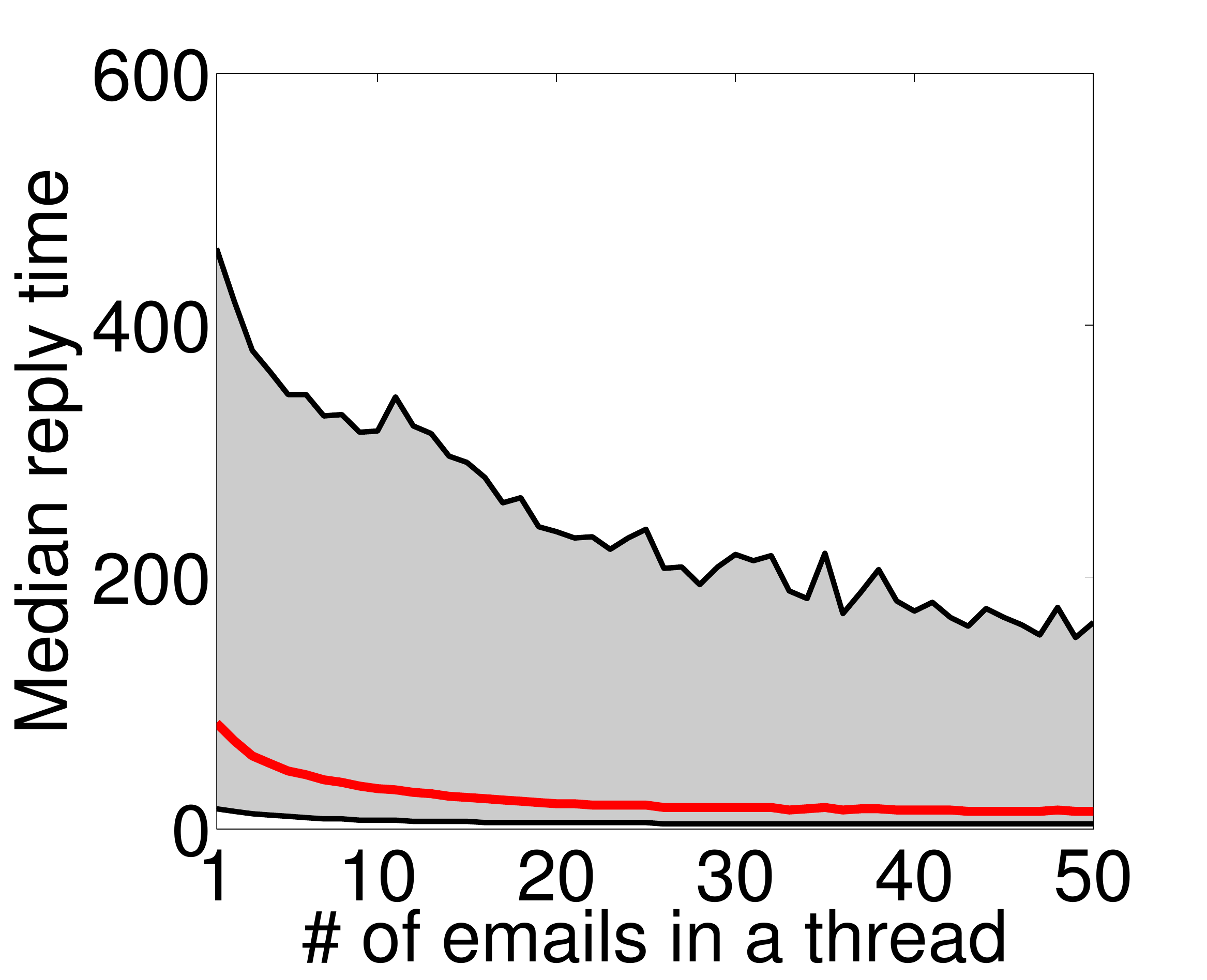}
      \label{fig:thread_time_corr}
   }
  &
   \subfigure[Length of reply]{
   \includegraphics[width=0.5\columnwidth]{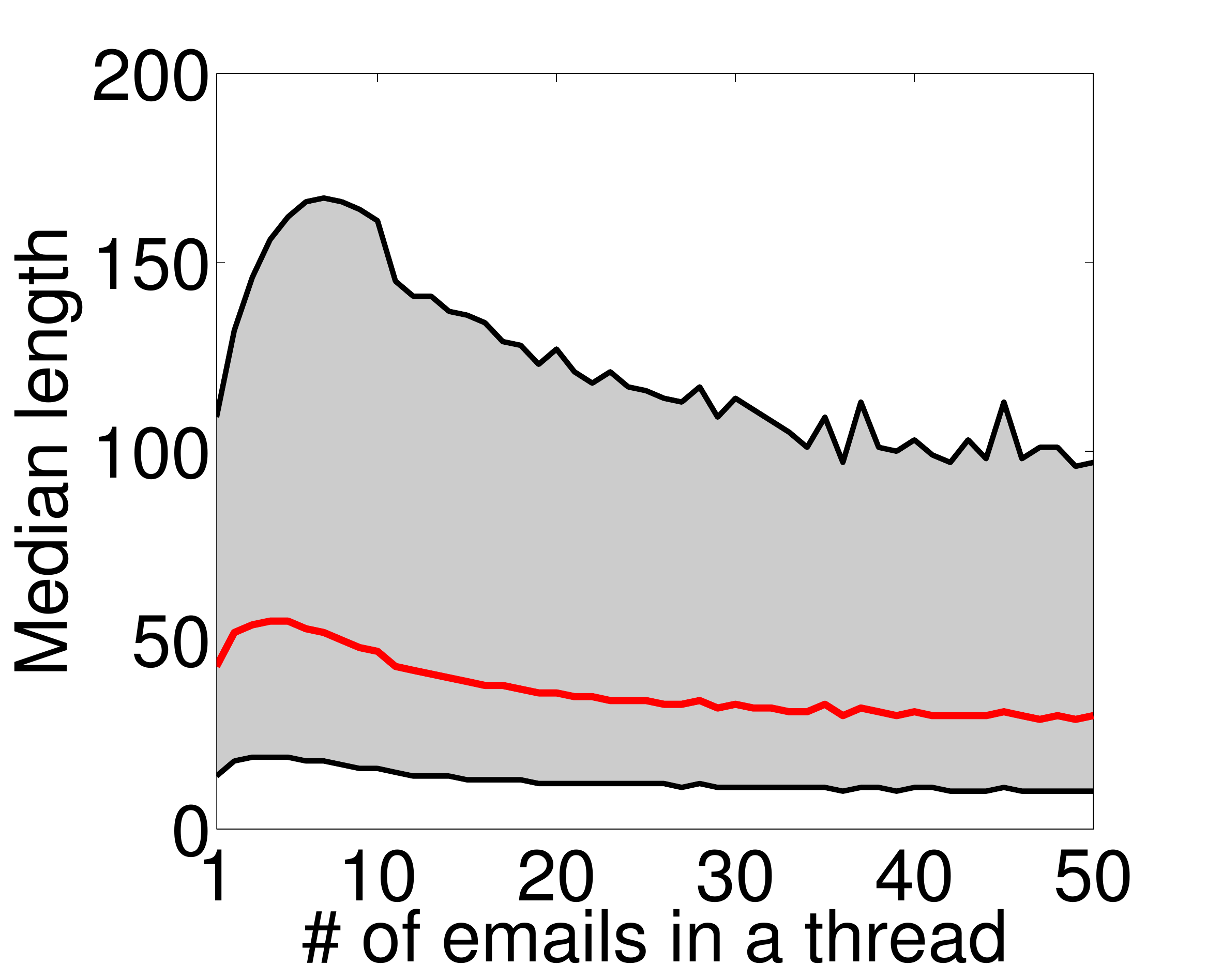}
   \label{fig:thread_length_corr}
   }
  \end{tabular}
   \caption{Reply time and length as a function of the length of a conversation for dyadic interactions with less than 50 steps in a thread, which are 99.7\% of all threads.  Each plot shows the median, 25$^{th}$ and 75$^{th}$ percentile of the measure vs. the number of messages in a thread. Longer threads have shorter reply delays and lengths. }
  \label{fig:time_length_thread_corr}
   \vspace*{-2mm}
\end{figure}

\noindent Next, we investigate the effect of the position of the reply within a conversation (i.e., step in a thread) on the reply time and length. Figure~\ref{fig:time_step} shows how reply time changes as a function of thread step, for threads of different length. Replies become faster as the conversation progresses, but the last reply is much slower than the previous replies. The long delay in a reply could be considered as a signal for the end of the conversation. Figure~\ref{fig:length_step} shows the effect of thread step on the length of the reply. Replies get slightly longer as conversation progresses, although the last reply in a thread is much shorter than the previous replies. Moreover, we see that longer conversations (threads) have faster reply times and shorter reply lengths. To better quantify this effect, we calculate the median reply time and length for different thread lengths. We find that both reply times and length of replies are smaller in longer conversations (Figure~\ref{fig:time_length_thread_corr}). This would be expected if the data covered a very short time period, because the reply times would had to be small to fit a long conversation in a short period of time. But, this is not the case here, since we are covering several months.

\begin{figure}[t!]
\begin{center}
\includegraphics[width=0.7\columnwidth]{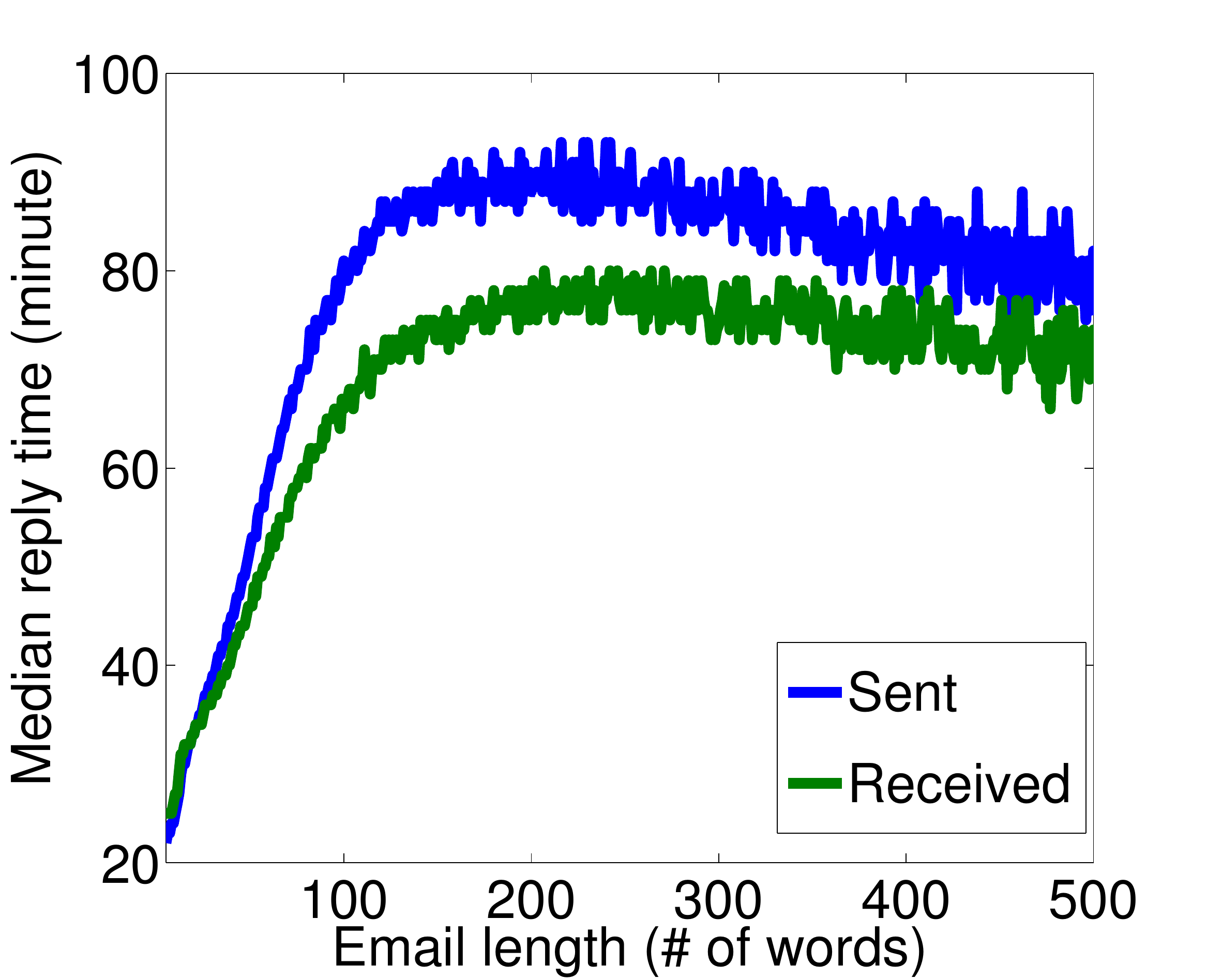}
\end{center}
\vspace*{-1mm}
\caption{Correlation between time to reply and length of reply for outgoing and incoming emails for dyadic conversations. There is a strong correlation till the length of 200 words (more than 83\% of all replies).}
\label{fig:length_time}
\vspace*{-3mm}
\end{figure}

Figure~\ref{fig:length_time} shows how time to reply to an incoming (received) message varies as the function of the length of the received message and the length of the reply.
There is a strong correlation between reply time and length, showing that longer replies take longer to be composed. However, replies longer than 200 words are slightly faster. This could be due to a number of reasons. First, we may not properly account for message length due to copy and pasted emails or missed quoted messages. Second, there could be systematic differences in the population of users who write replies longer than 200 words, e.g., such users may be more adept at writing.
There is also a strong correlation between reply time and the length of received emails, showing that the longer the messages the users receive, the longer it takes them to reply. The slight decrease in reply time for messages longer than 200 words could be explained as above.

\section{Factors Affecting Replying}\label{section:factors}
\noindent A variety of factors affect replying behavior, most obviously the contents of the messages and who they are from. These are idiosyncratic and highly variable. Instead, in this section we examine universal factors that affect reply time and length, such as the time of the day the message was received, user's demographic characteristics, and volume of emails received. In Section~\ref{section:prediction}, we show that these factors help predict user replying behavior even when email content is not known.
 

\subsection{Circadian Rhythms}
\begin{figure}[t!]
\begin{tabular}{@{}c@{}c@{}}
\subfigure[Reply time]{
   \includegraphics[width=0.5\columnwidth]{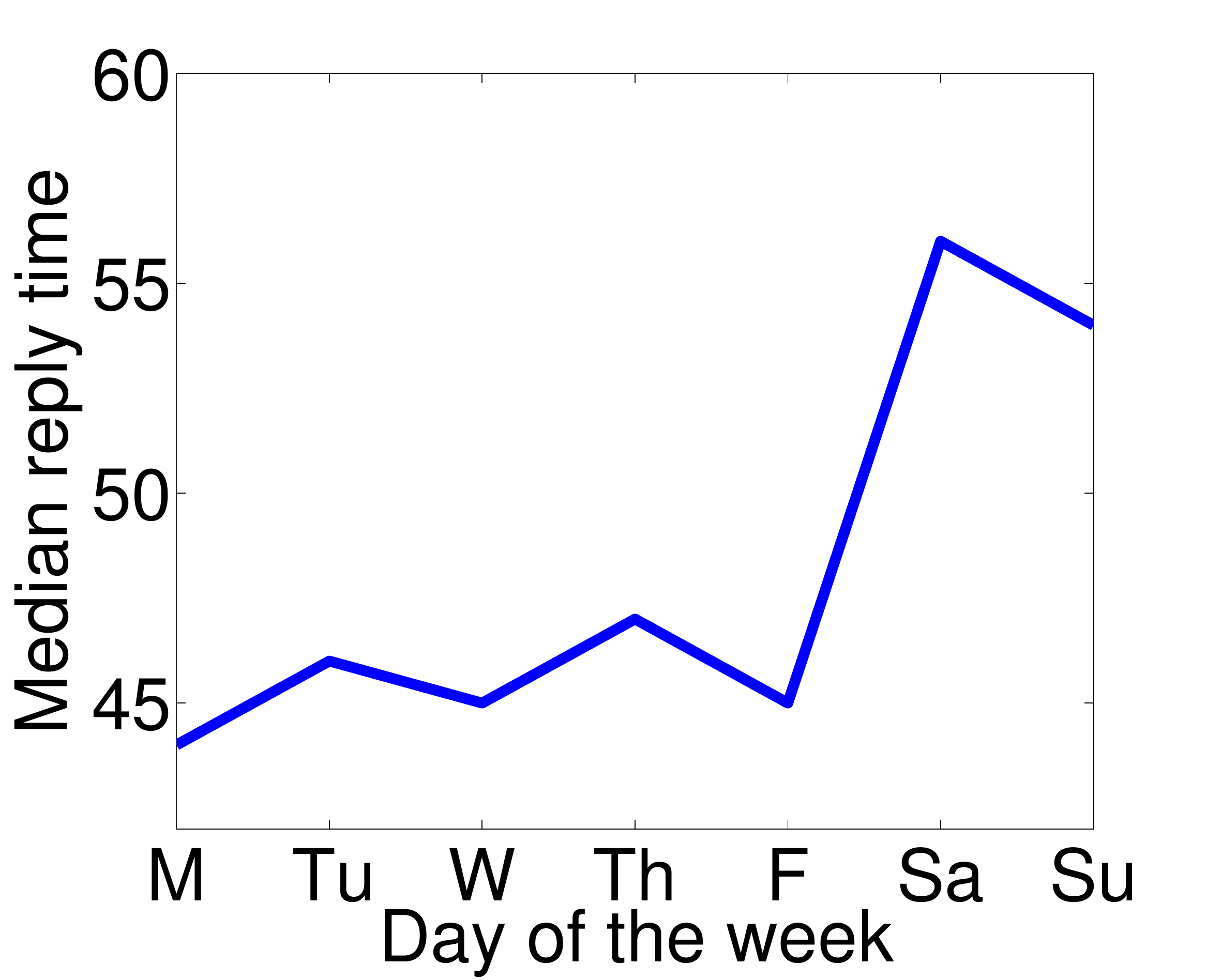}
      \label{fig:day_time}
   }
  &
   \subfigure[Reply length]{
   \includegraphics[width=0.5\columnwidth]{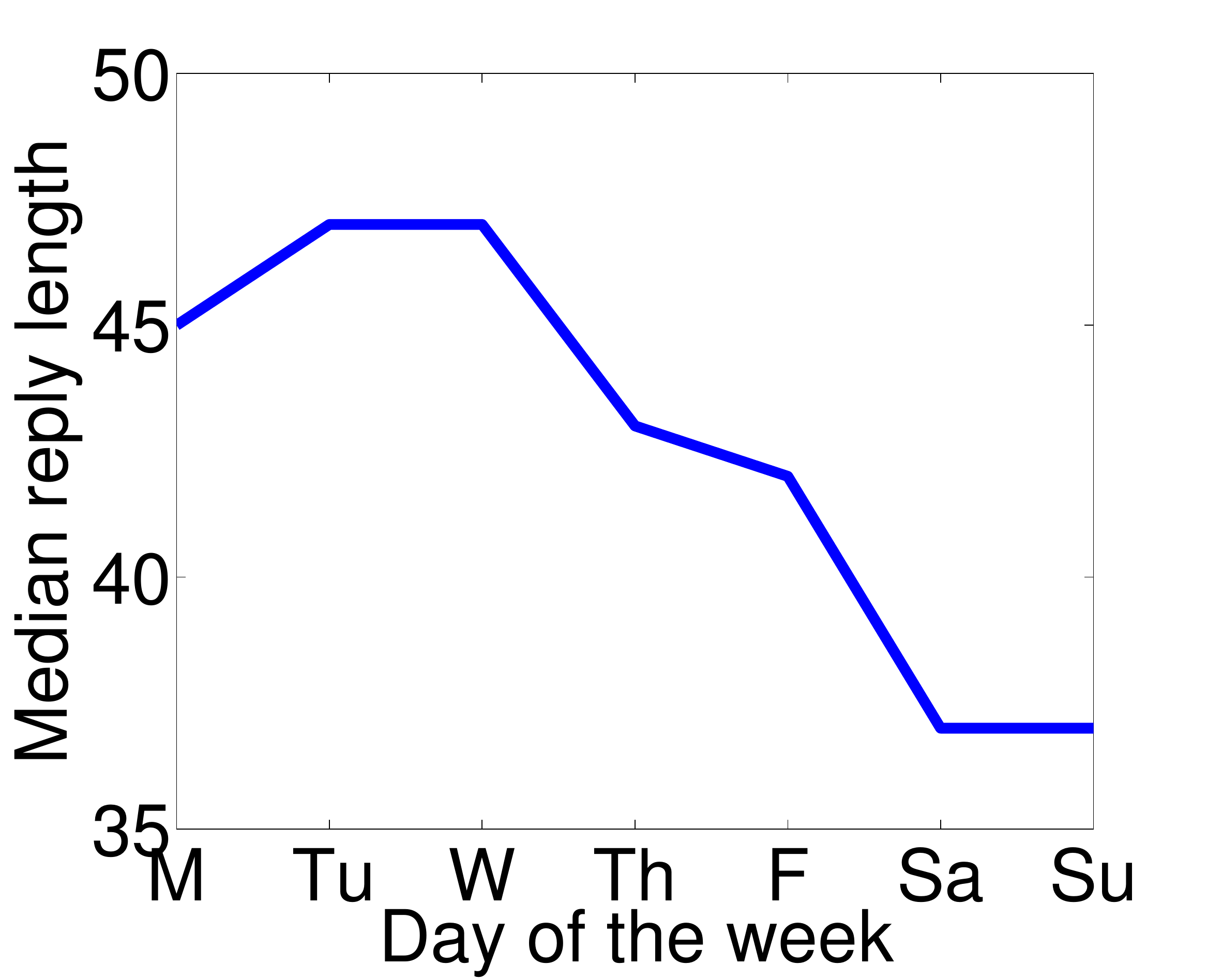}
   \label{fig:day_length}
   } \\
   \subfigure[Reply time]{
   \includegraphics[width=0.5\columnwidth]{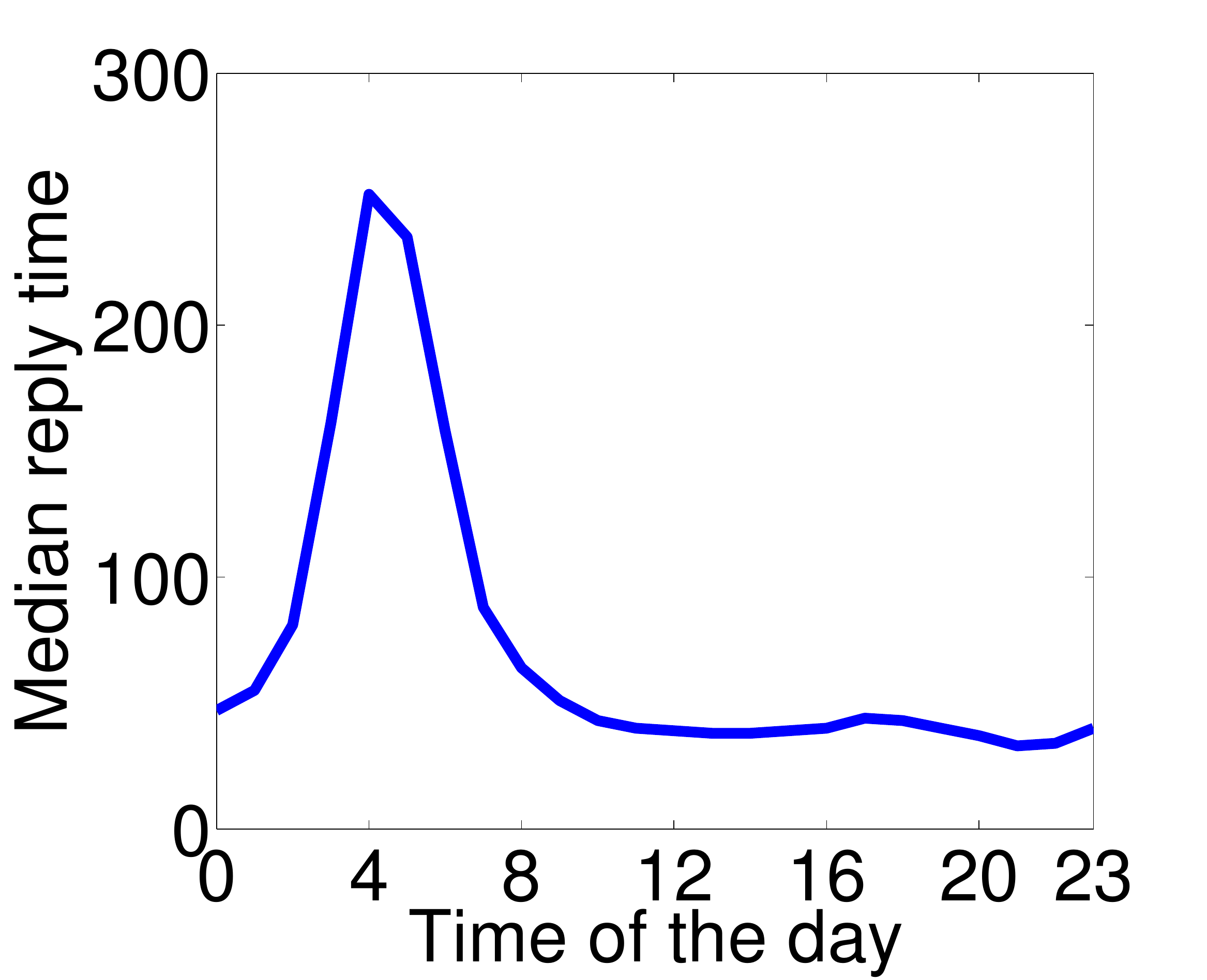}
      \label{fig:hour_time}
   }
  &
   \subfigure[Reply length]{
   \includegraphics[width=0.5\columnwidth]{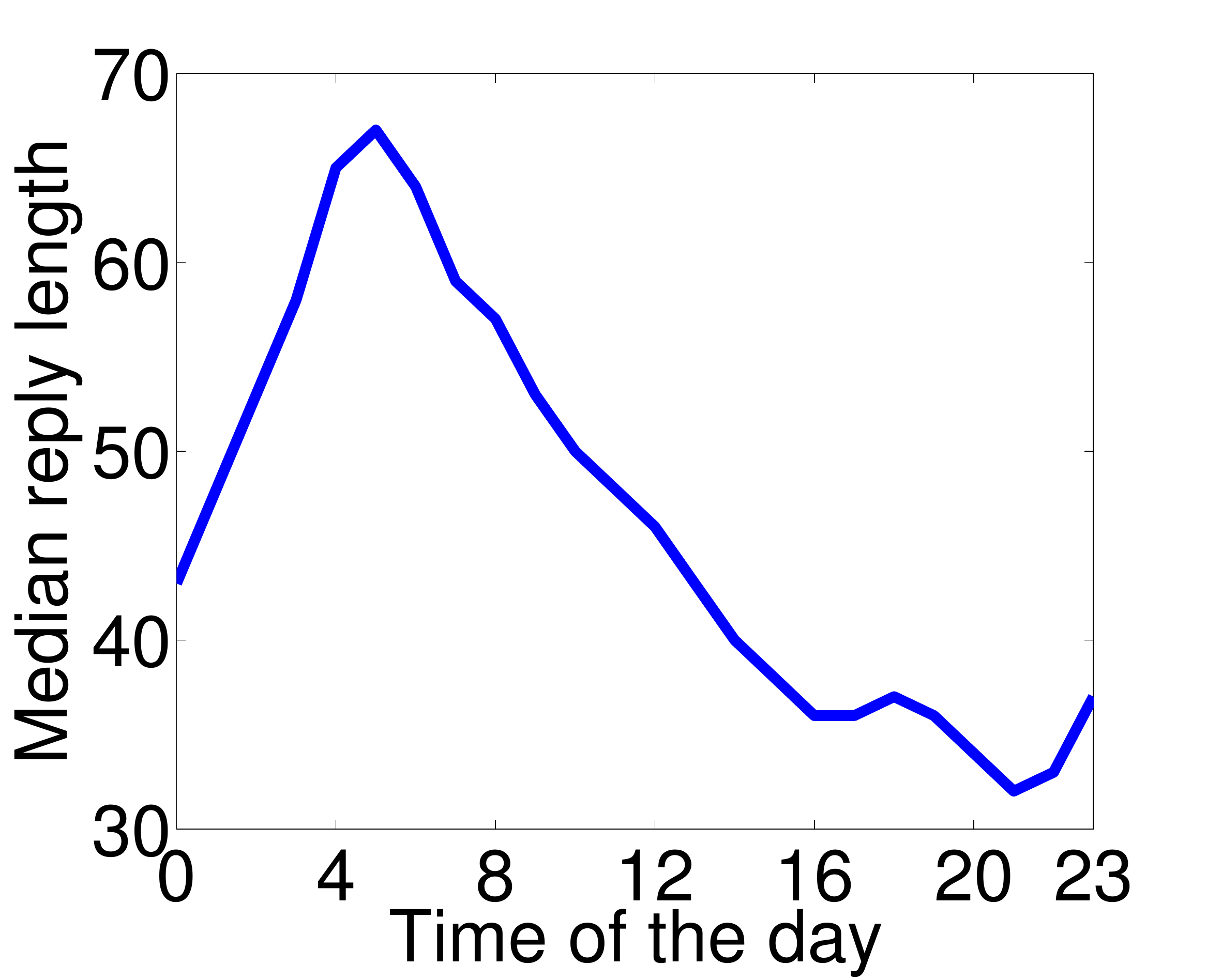}
   \label{fig:hour_length}
   }
  \end{tabular}
   \caption{Day of the week and time of the day effects on email reply time and length. (a) Reply time in minutes and (b) length in number of words as a function of  day of the week the message replied to was received. Emails received on weekends get shorter and slower replies, compared to the workdays. (c) Reply time and (d) length as a function of the hour of the day the email was received. Emails received during the night have considerably longer reply times.}
  \label{fig:day_time_length}
  \vspace{-2mm}
\end{figure}

\noindent Email users are more active during the day than night time, and also on workdays rather than the weekend. These circadian cycles have a strong effect on email replying behavior.
Figure~\ref{fig:day_time_length} shows how the day of the week and the hour of the day the message was received, accounting for the time-zones, affect how quickly the email is replied to and the length of the reply. Emails received on weekends get substantially shorter replies compared to those received on workdays (Fig.~\ref{fig:day_length}). The replies are also much slower (Fig.~\ref{fig:day_time}). Messages received during the night get slower replies than those received during working hours (Fig.~\ref{fig:hour_time}). Interestingly, messages received in the morning get substantially longer replies than those received in the afternoon and evening (Fig.~\ref{fig:hour_length}).

\subsection{Demographics of Users}

\begin{figure}[thb!]
\begin{tabular}{@{}c@{}c@{}}
\subfigure[Age and reply time]{
   \includegraphics[width=0.5\columnwidth]{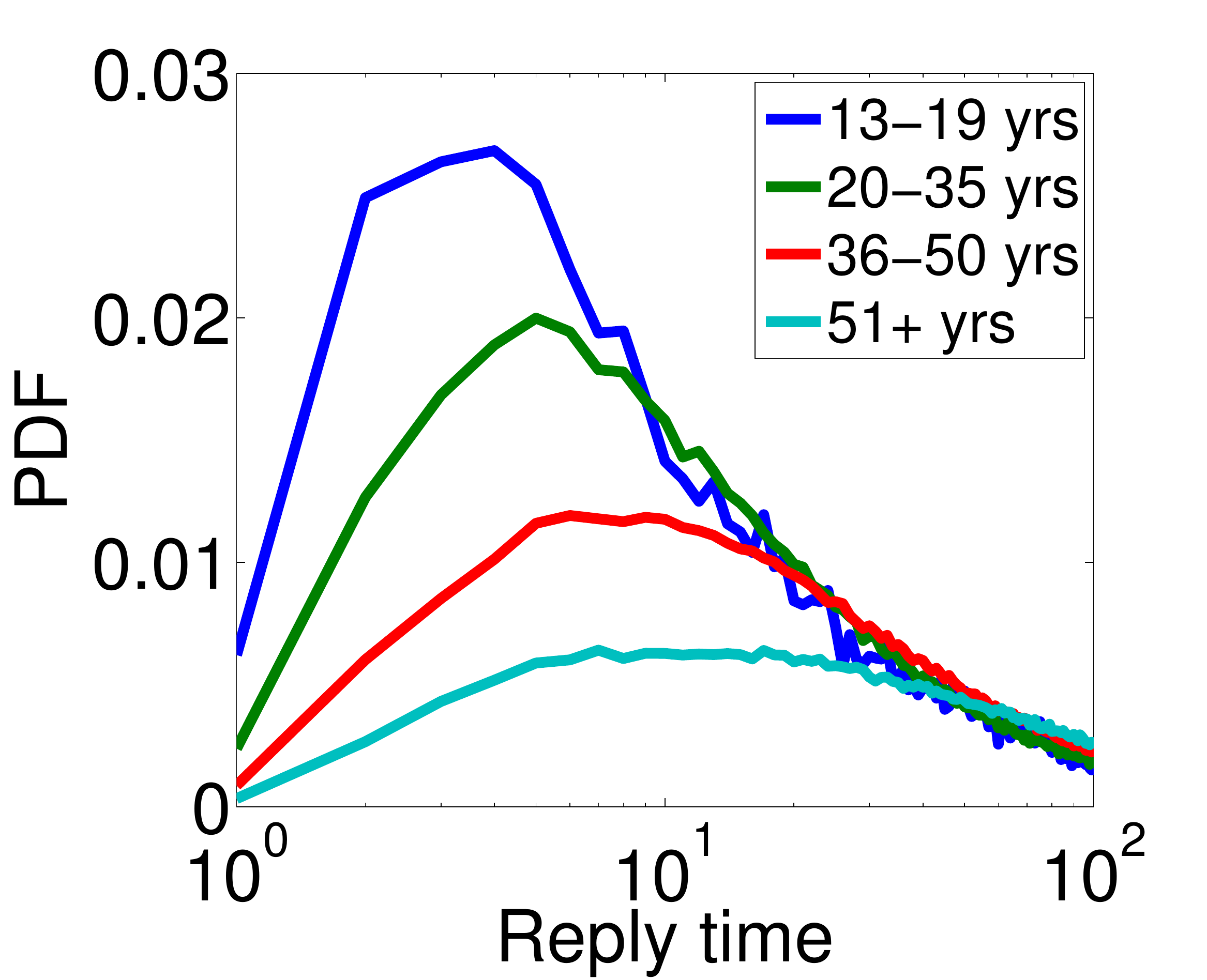}
      \label{fig:time_age}
   }
  &
   \subfigure[Gender and reply time]{
   \includegraphics[width=0.5\columnwidth]{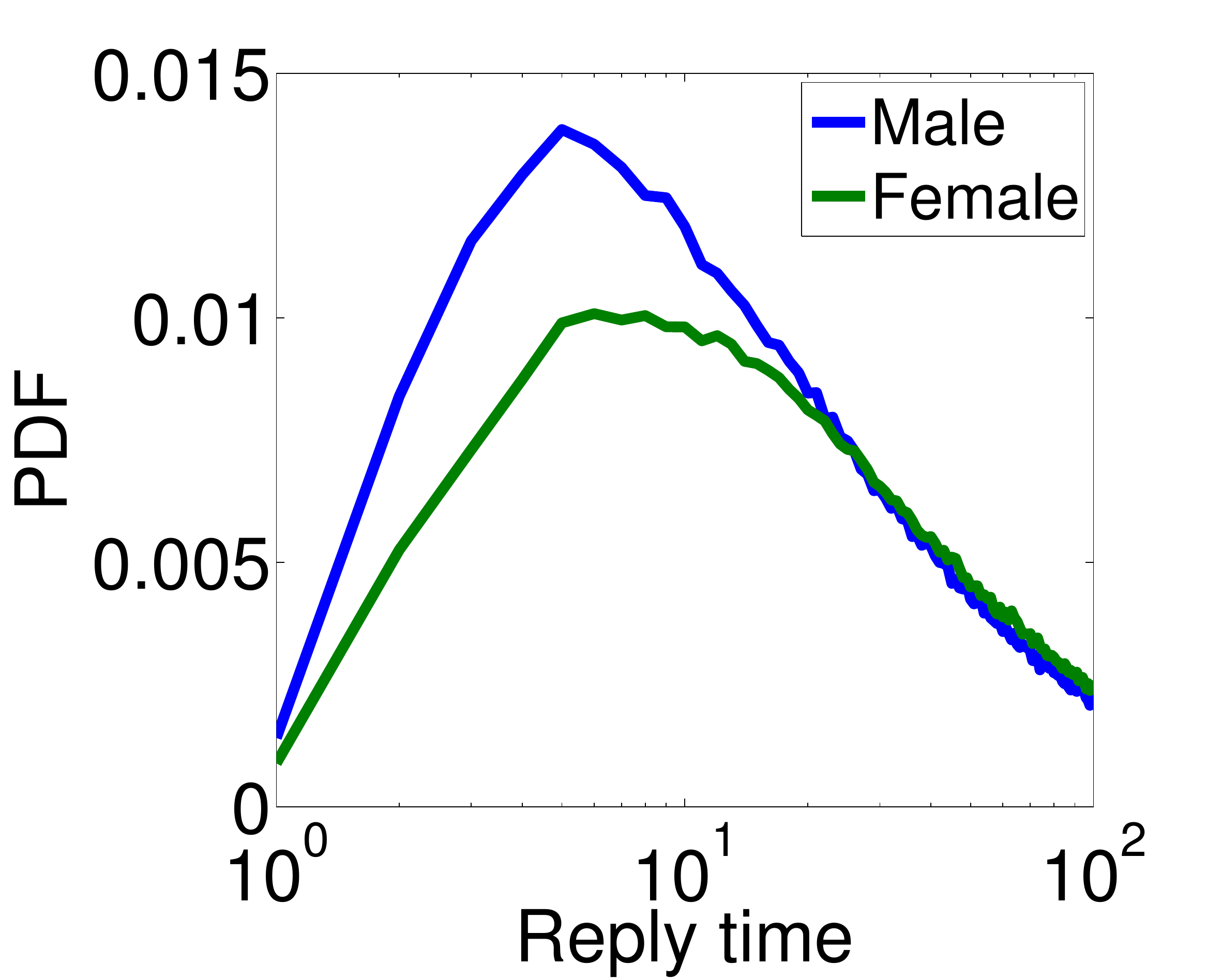}
   \label{fig:time_gender}
   }\\
   \subfigure[Age and reply length]{
   \includegraphics[width=0.5\columnwidth]{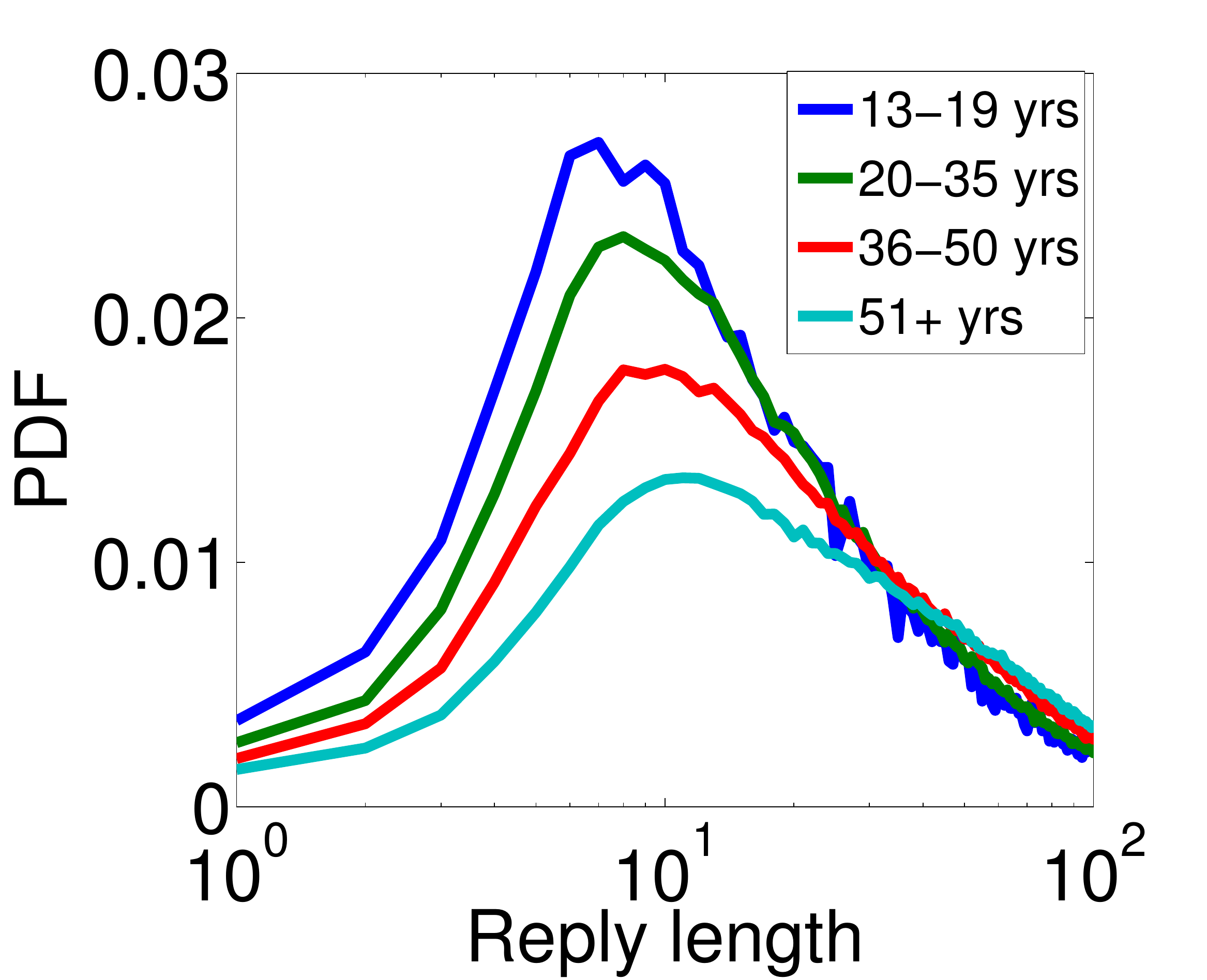}
      \label{fig:length_age}
   }
  &
   \subfigure[Gender and reply length]{
   \includegraphics[width=0.5\columnwidth]{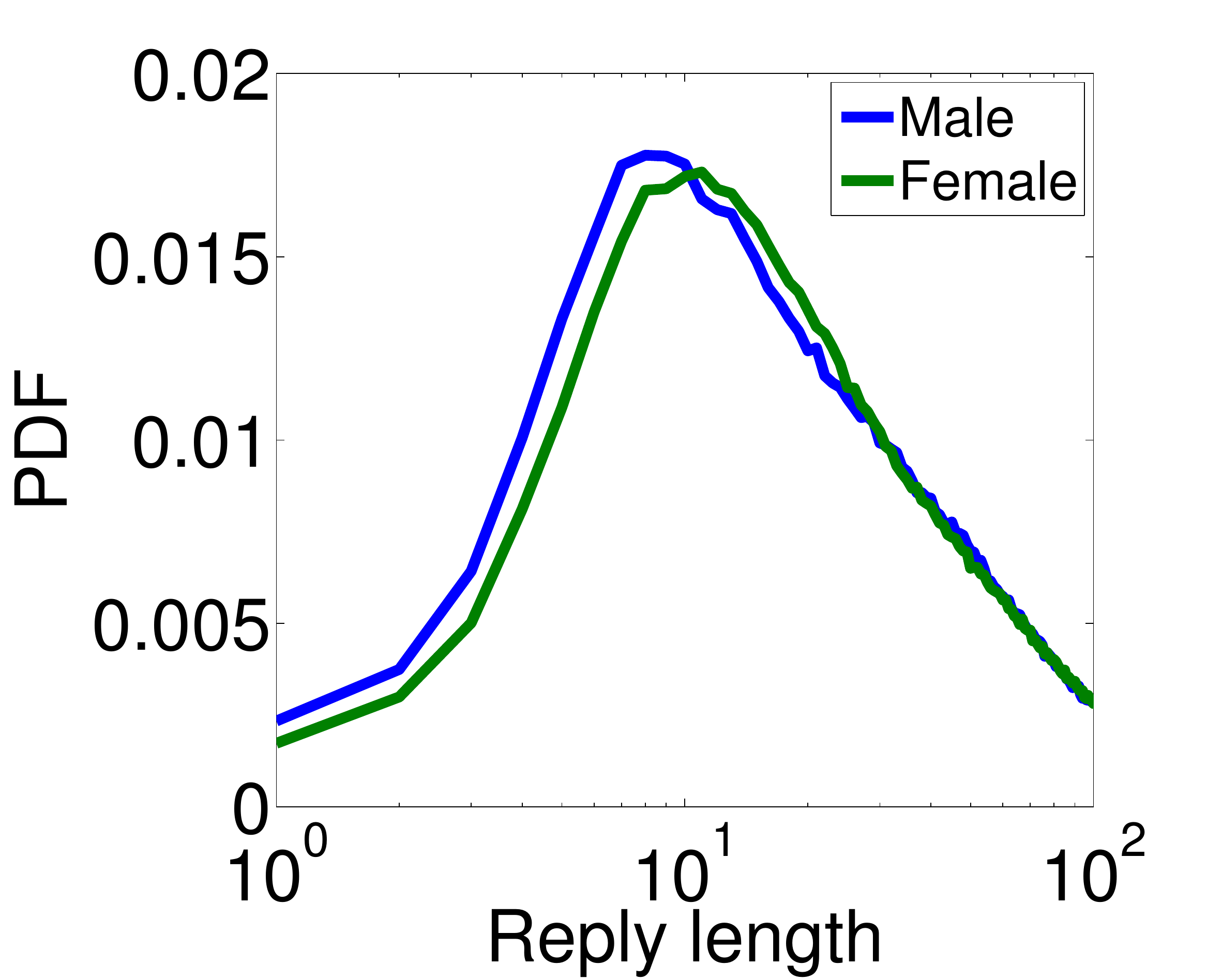}
   \label{fig:length_gender}
   }
  \end{tabular}
   \caption{Demographics and email replying behaviors. PDF of reply time of users categorized by (a) age and (b) gender. Teens are fastest to reply, and as users get older their replies become slower. PDF of reply length of user populations divided by (c) age and (d) gender. Teenagers have the shortest replies, and as users get older their replies become longer. }
  \label{fig:time_demo}
\end{figure}

\begin{figure}[thb!]
\begin{tabular}{@{}c@{}c@{}}
\subfigure[Age]{
   \includegraphics[width=0.5\columnwidth]{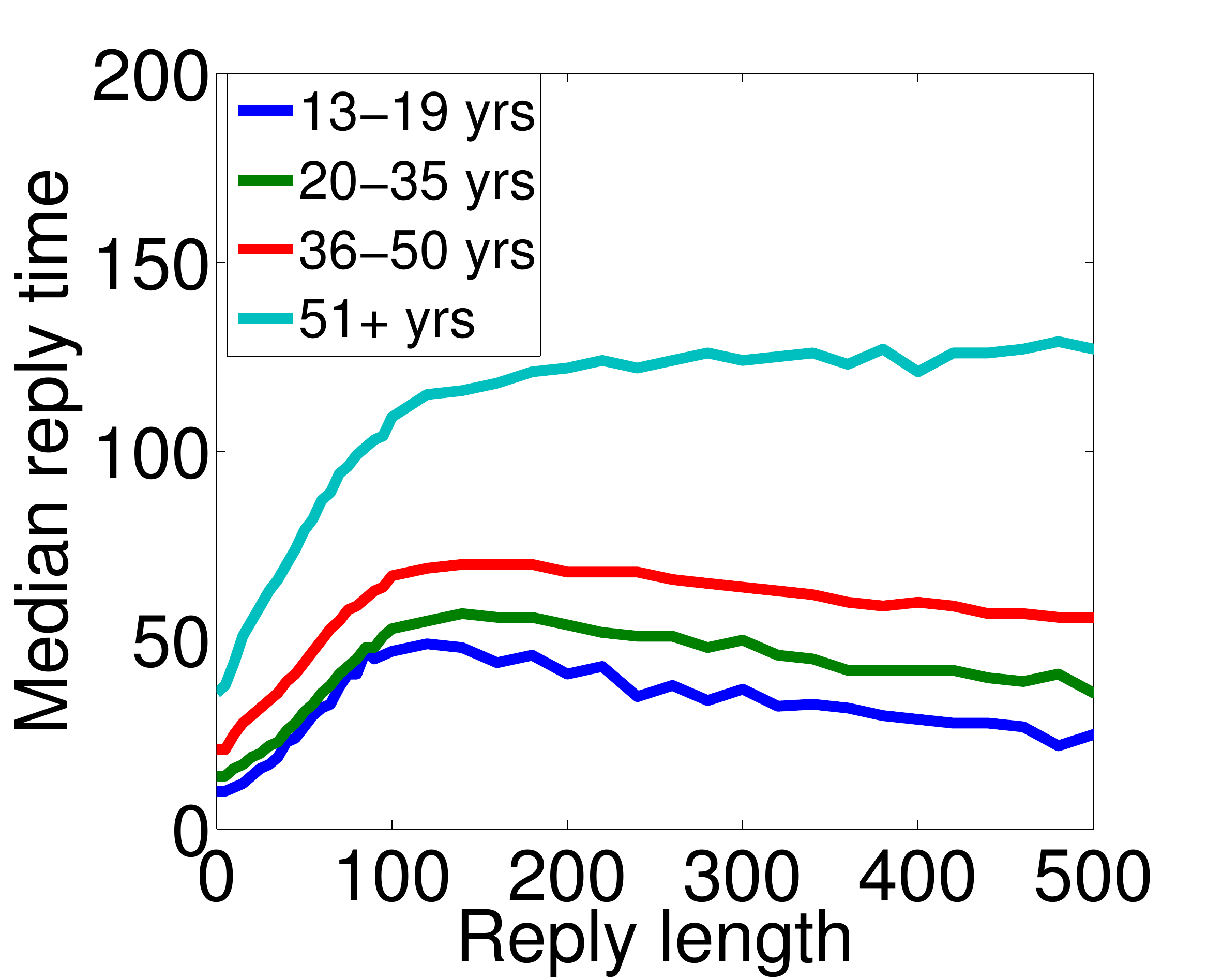}
      \label{fig:length_time_trend_age}
   }
  &
   \subfigure[Gender]{
   \includegraphics[width=0.5\columnwidth]{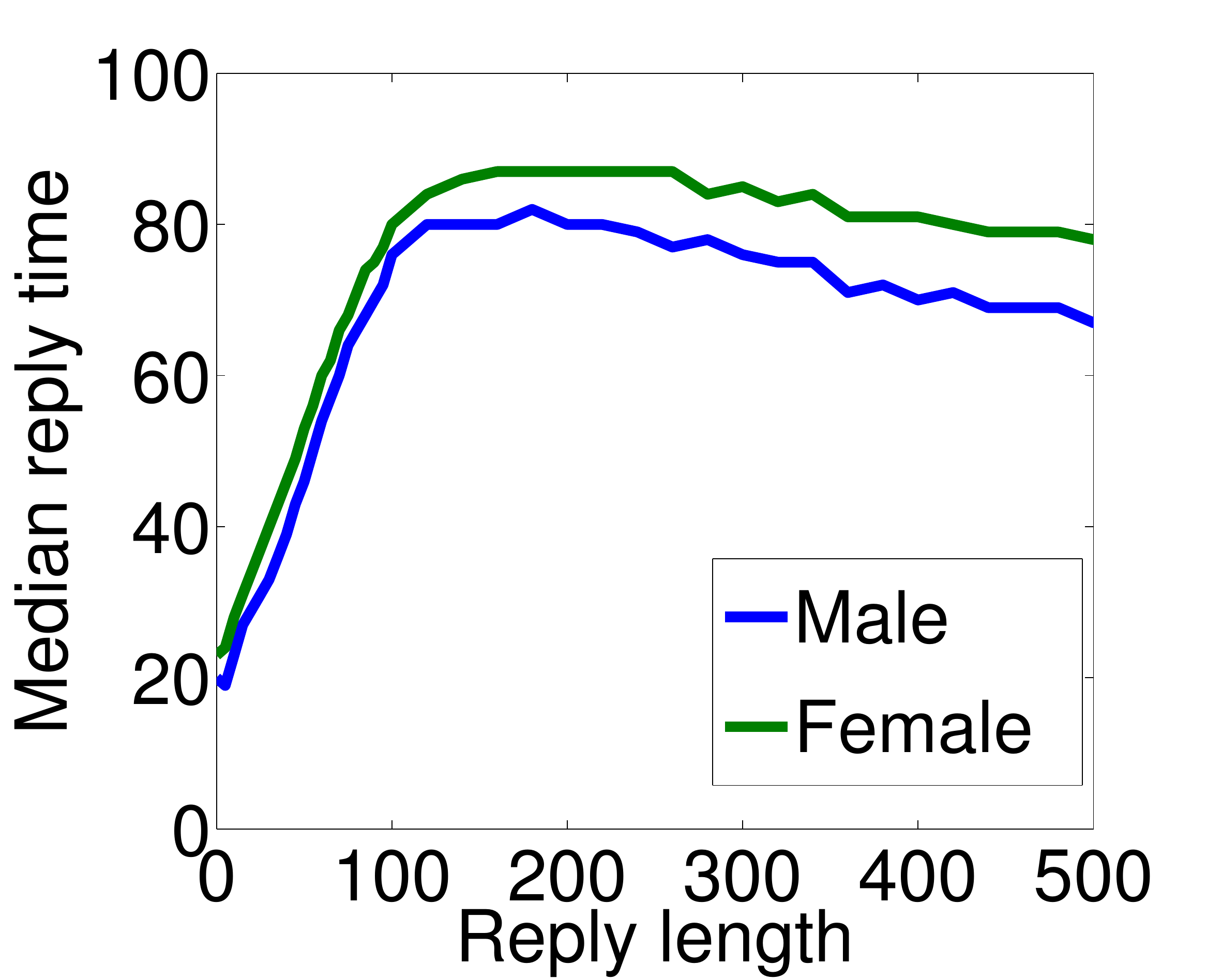}
   \label{fig:length_time_trend_gender}
   }
  \end{tabular}
   \caption{Median time to reply, given the length of the reply. Younger users are faster in composing the emails with the same length, compared to older users, and males are slightly faster than females.}
  \label{fig:length_time_demo}
\end{figure}

\noindent Next, we investigate how demographic factors affect replying behavior.
We categorize users based on their age and gender and compare email replying behavior across different populations. Results of demographic analysis of reply time are presented in Figure~\ref{fig:time_demo}. Youngest email users, teens, have the fastest reply times; as users get older they become slower to reply to emails. Median reply time of different age groups was as follows: 13 minutes for teens, 16 minutes for young adults (20--35 years old), 24 minutes for adults (36--50 years old), and 47 minutes for mature users (51 and older). Gender does not seem to play as important a role in the replying behavior: females and males have very similar distributions of reply time, with men being slightly faster, with median reply time of 24 minutes, compared to 28 minutes for women.

The length of replies is also affected by demographics. Similar to reply time distribution, user groups who send faster replies, also send shorter replies: teens send the shortest replies and older users send longer ones. Teenagers' median reply length is only 17 words, while young adults' median reply length is 21 words, for adults it is 31 words, and for mature users it is 40 words. There is again no considerable difference between males and females, males having median length of 28 words, compared to females with median length of 30 words.

The older users are in general sending replies with longer delays, but they are also sending longer replies, which might explain their longer delays. To test if the difference in reply time is caused by the difference in length of replies, we compare the reply time of the users with different ages, while we account for the length of the reply. When plotting the median reply time for particular length of replies (Figure~\ref{fig:length_time_demo}) the difference between young and older users becomes even clearer. Teenagers typically spend 39 minutes for sending a reply with 50 words, whereas older users (51+ years old) spend 79 minutes for the replies with the same length. Interestingly, as the length of the reply increases, the gap between the older users and younger users also increases. We repeat the same analysis for the gender of the users and the males are slightly faster than the females in composing replies of the same length.

\subsection{Mobile Devices}
\noindent We did not have access to email metadata, which had such useful information as the email client and the device used to send the email. However, we were able to reconstruct some of this information, because many mobile devices add a signatures, like ``Sent from my iPhone'', to the sent emails by default. Using such signatures, we were able to identify emails that were sent from smart phones or tablets and compare them with the emails from desktop computers. While this procedure does not detect all emails from mobile devices, the large size of our corpus gives us enough data to look for systematic differences in the replying behavior of mobile device users. Hence, to categorized emails as being sent from a phone, a tablet, or a desktop,  we search for signatures such as ``Sent from my iPhone'' and ``Sent from my iPad'' (among others), after removing the quoted message, so that we would not catch a signature in the quoted message.

We find that replies sent from phones are the fastest, followed by emails sent from tablets, and finally replies from desktops. Emails from phones have a median reply time of only 28 minutes, compared to the 57 minutes for the replies from tablets, and 62 minutes for the desktop. This is expected since users often configure their phones to notify them of a received email, and they usually have their phones with them. Also, replies sent from mobile devices tend to be shorter than those sent from desktops. Replies from phones have a median length of 20 words, and tablets have a median length of 27 words, while desktops have median of 60 words. 

Interestingly,  adults (35--50 years old) represent the highest percentage of mobile device users, with 53\% of them using a phone or a tablet at least once. Teens and young adults are the next largest populations of mobile users, with 49\% and 48\%, respectively. Mature adults use mobile devices the least: only 43\% of them do. Also, more women use mobile devices than men (50\% vs. 45\%).

\subsection{Attachments}

\noindent We also study the effect of attachments in received emails on the time and length of replies. Replies to emails with attachments are much slower (median of 56 minutes) than replies to emails without any attachment (median of 32 minutes). The difference of almost a factor of two could be explained by the time needed for reading the attachment. Also, the emails with an attachment, get longer replies (median of 47 words) than emails without attachments (median of 33 words), which is probably due to the fact that the replier has to respond to more information.

\subsection{Email Overload}
\noindent How does replying behavior change as the number of incoming emails increases? In other words, as users receive more emails in a day, how do they adapt to the increased information load: do they become more active and reply to more emails, do they delay replying or fail to reply altogether?

We characterize a user's \emph{email load} by the number of messages the user receives in a day, and the user \emph{activity} by the number of sent emails.  We divide users into two groups based on their activity: \textit{low activity} and \textit{high activity}. To do this, we rank users by their activity and consider the top third to be the \textit{high activity} users and the bottom third as \textit{low activity} users.


\begin{figure}[t!]
\begin{center}
\includegraphics[width=0.8\columnwidth]{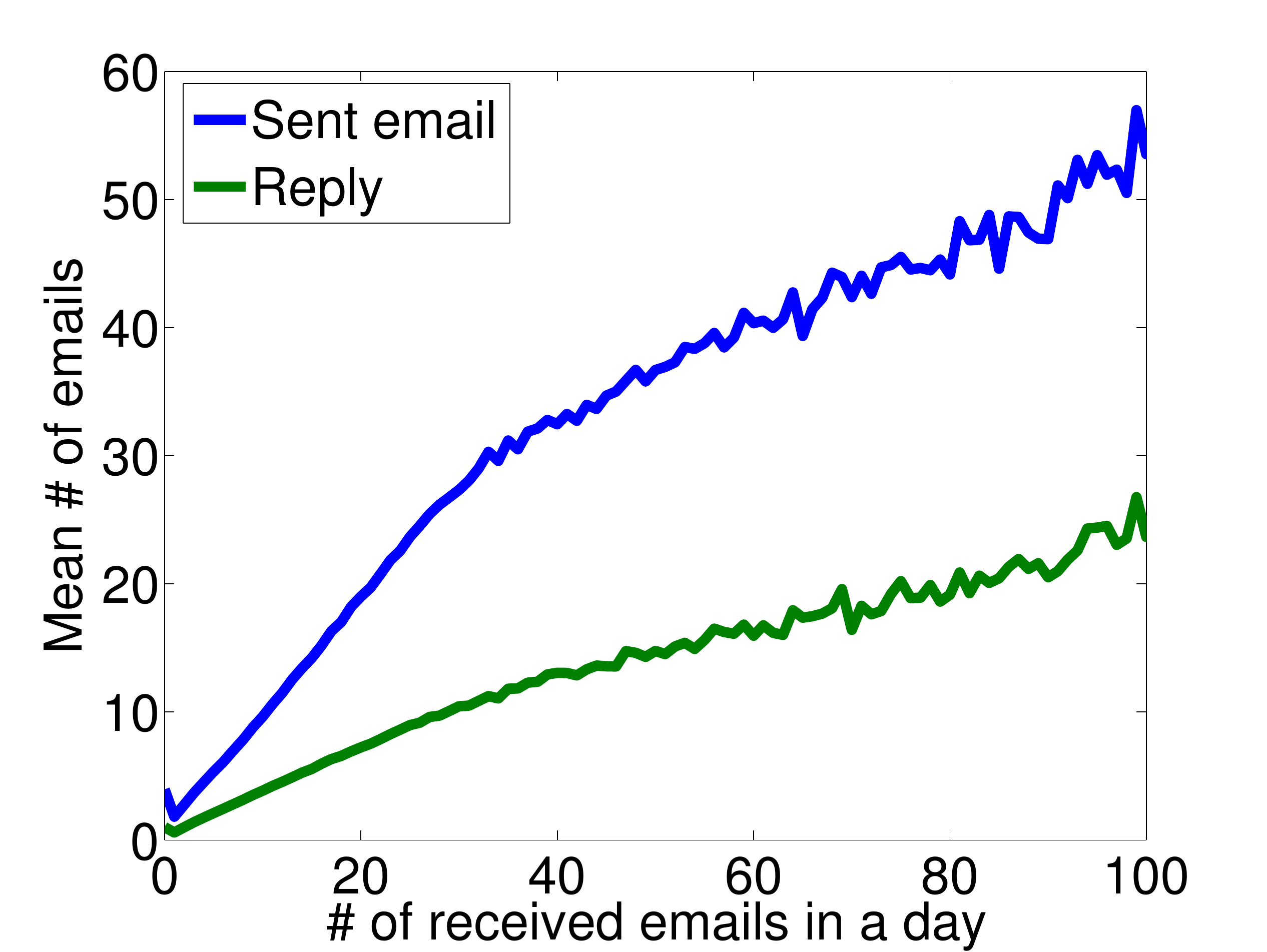}
\end{center}
\caption{Change in behavior due to increased email load. Users send and reply to more emails, as they receive more emails in a day.}
\label{fig:sent_reply_load}
\end{figure}

Email load affects user behavior. Users increase their activity as their email load, i.e., the number of emails received that day, grows. Figure~\ref{fig:sent_reply_load} shows that user activity increases, both in terms of the number of sent emails and replied emails, as the number of emails they receive in a day grows. Here, we eliminate the users who sent more than 1,000 emails in a day, which is equivalent to sending more than one email in a minute for 16 hours straight. Such high activity is likely generated by bots, rather than humans.

\begin{figure}[t!]
\begin{tabular}{@{}c@{}c@{}}
\subfigure[Fraction of replies]{
   \includegraphics[width=0.5\columnwidth]{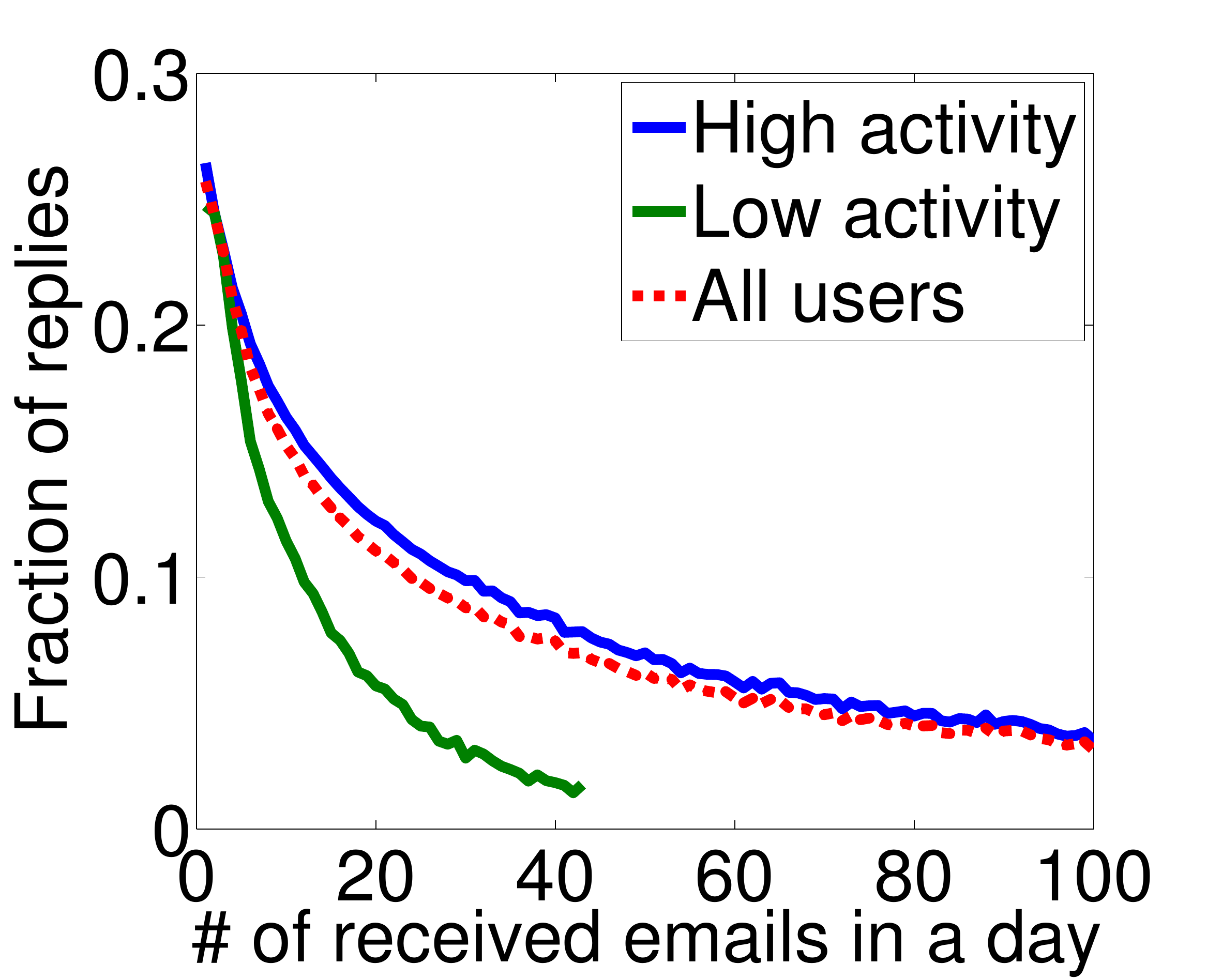}
      \label{fig:reply_frac_2group}
   }
  &
   \subfigure[Fraction of replies to email from contacts]{
   \includegraphics[width=0.5\columnwidth]{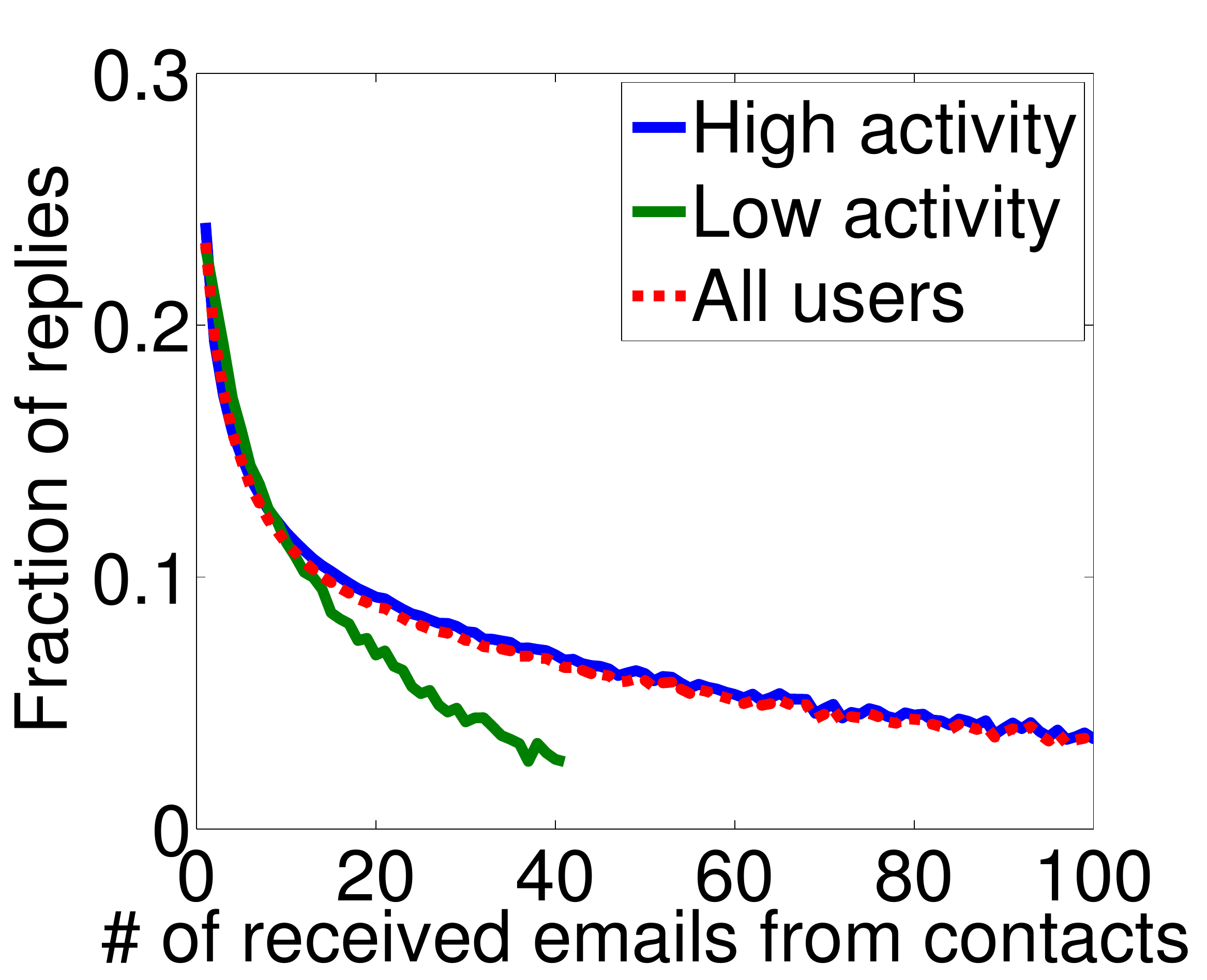}
   \label{fig:reply_frac_contact_2group}
   }
  \end{tabular}
   \caption{Fraction of replies given the number of emails received in a day. (a) The number of received emails includes all emails. (b) Only emails received from contacts who the user had sent a message to are counted as received emails.}
  \label{fig:reply_frac_2}
  \vspace{-3mm}
\end{figure}

While users increase their activity with higher email load, it appears that they are not able to adequately compensate for the increased load. Figure~\ref{fig:reply_frac_2group} shows that 
as the email load increases, users reply to a smaller fraction of their emails, from about 25\% of all emails received in a day at low load to less than 5\% of emails at high load (about 100 emails a day). However, highly active users are better able to keep up with the rising email load than low activity users.


Of course, not all emails require replies: spam, mailing list, advertisements, and purchase notifications are generally not replied to. As the number of such emails a user receives increases, the user might still be able to keep up with interactions with contacts, so the decreasing fraction of replies may not signal email overload. To address this question, we restrict analysis to emails received from contacts only, i.e., Yahoo mail users who emailed each other in our data set. The overall trend shown in Figure~\ref{fig:reply_frac_contact_2group} is the same: as email load increases, users reply to an ever smaller fraction of emails. This suggests that information overload is a problem, and users are unable to keep pace with rising incoming email traffic. Note that we cannot compare figures ~\ref{fig:reply_frac_2group} and~\ref{fig:reply_frac_contact_2group} with each other, since a user with a particular load in Figure~\ref{fig:reply_frac_2group} would move to the left in Figure~\ref{fig:reply_frac_contact_2group} and might bring down the fraction for the days with low email load.

\begin{figure}[t!]
\begin{tabular}{@{}c@{}c@{}}
\subfigure[Age]{
   \includegraphics[width=0.5\columnwidth]{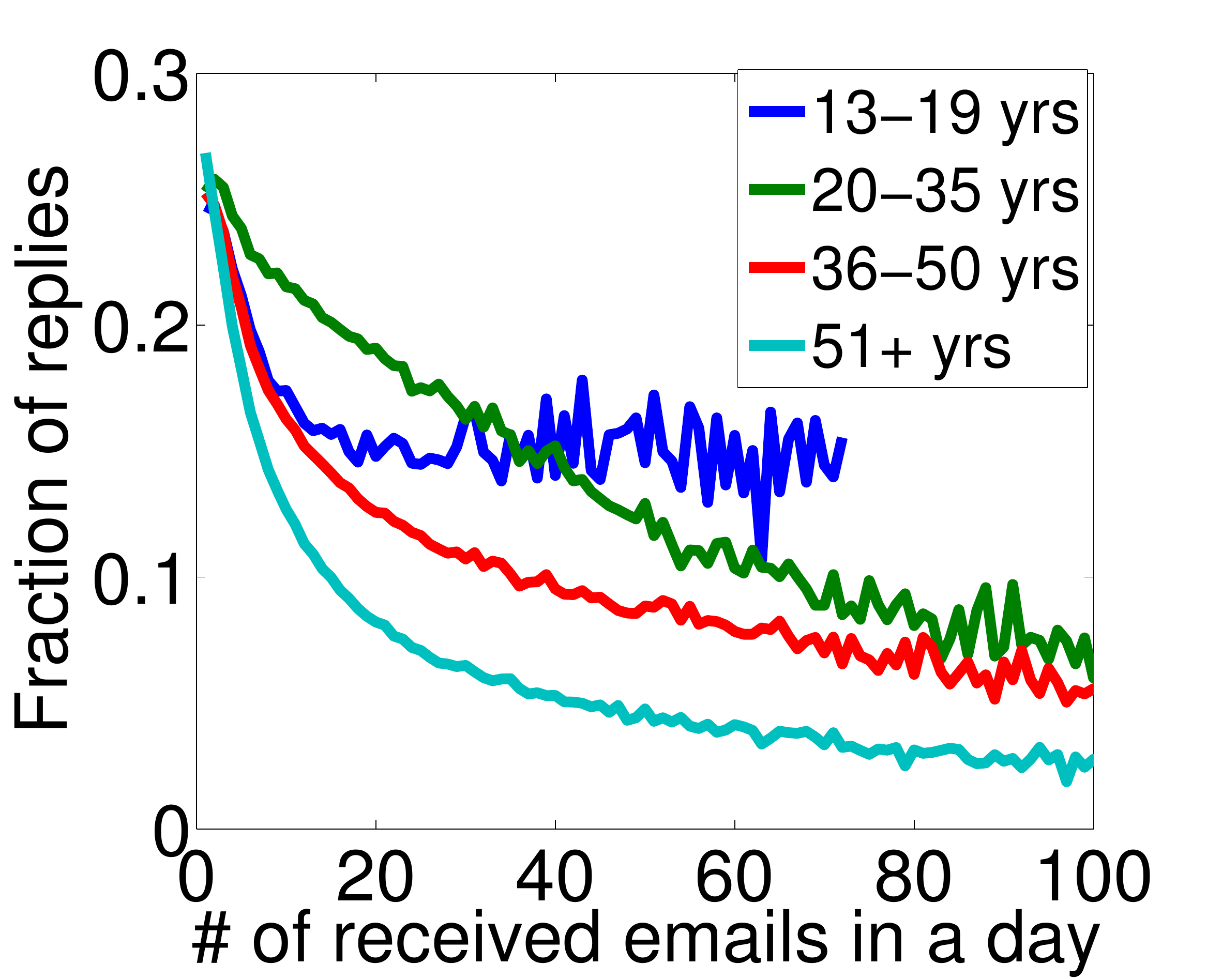}
      \label{fig:reply_frac_age}
   }
  &
   \subfigure[Gender]{
   \includegraphics[width=0.5\columnwidth]{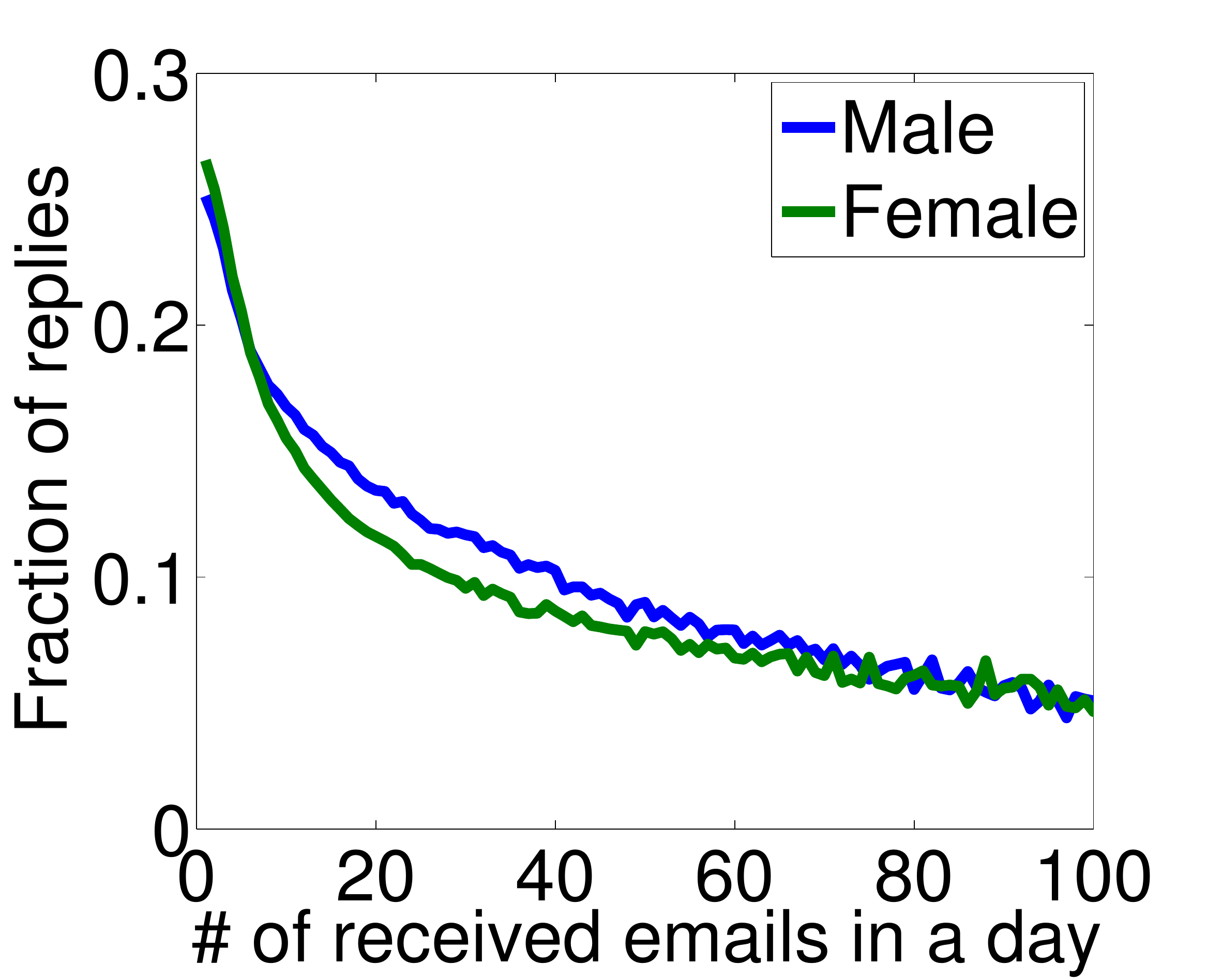}
   \label{fig:reply_frac_gender}
   }
  \end{tabular}
   \caption{Fraction of replies given the number of emails received in a day for different demographic populations. Younger users are less sensitive to email overload than older users, while men are similar to women in their response. Reply time is for emails from dyadic interactions, but received emails include all emails. }
  \label{fig:reply_frac_demo}
  \vspace{-3mm}
\end{figure}

Next, we categorize users based on their demographics to see how they respond to email overload. 
Figure~\ref{fig:reply_frac_age} shows the result for different age groups. Younger users can deal with the increasing information load much better than older users. Teens seem to experience little overload, replying to a constant fraction of emails, even as the load increases. Email overload becomes progressively worse for older age groups. There is little difference in how different genders respond to email load, with women slightly more affected by the increase in the load of information than men (Figure~\ref{fig:reply_frac_gender}).

\begin{figure}[t!]
\begin{tabular}{@{}c@{}c@{}}
\subfigure[Reply time]{
   \includegraphics[width=0.5\columnwidth]{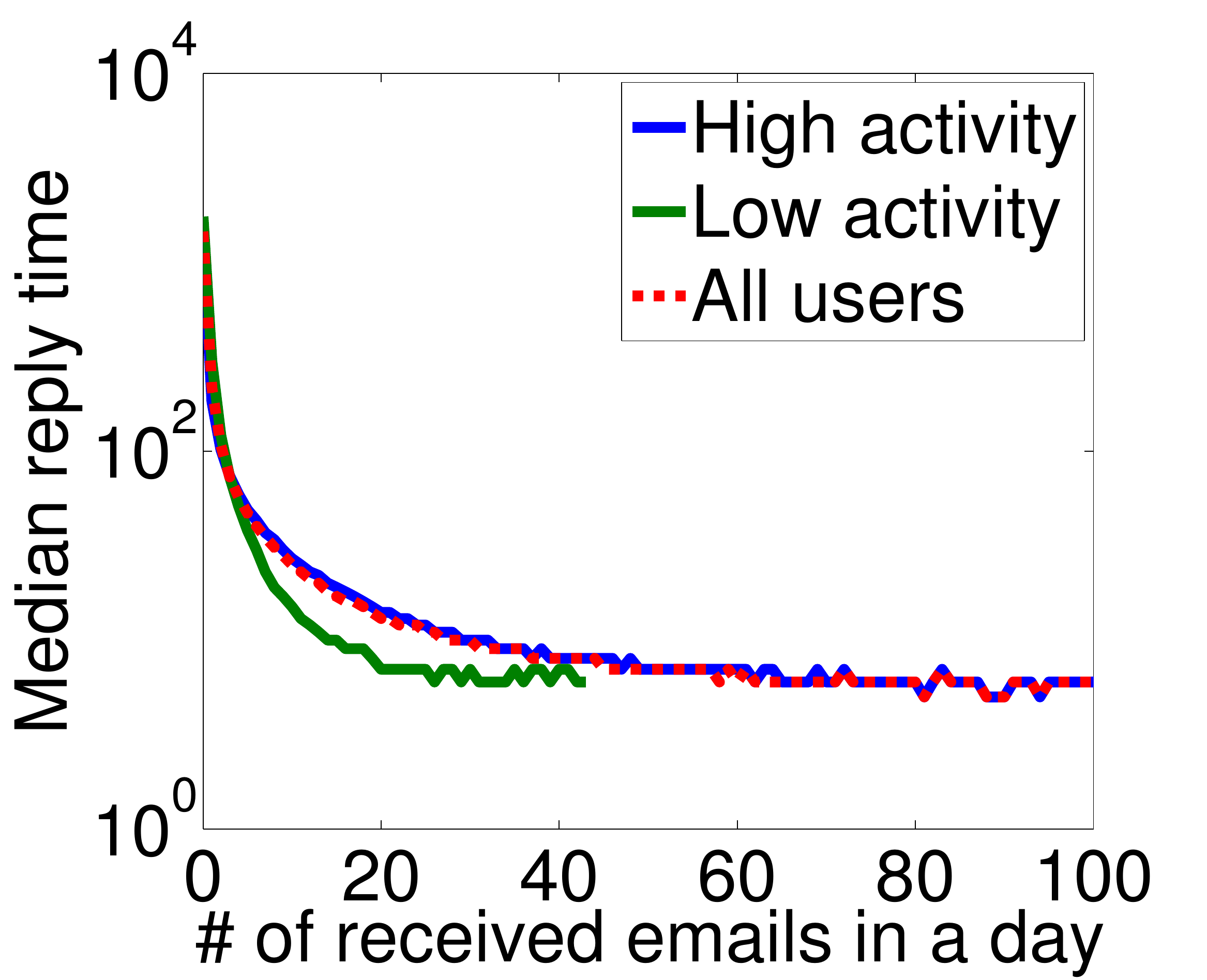}
      \label{fig:time_over_2group}
   }
  &
   \subfigure[Length of reply]{
   \includegraphics[width=0.5\columnwidth]{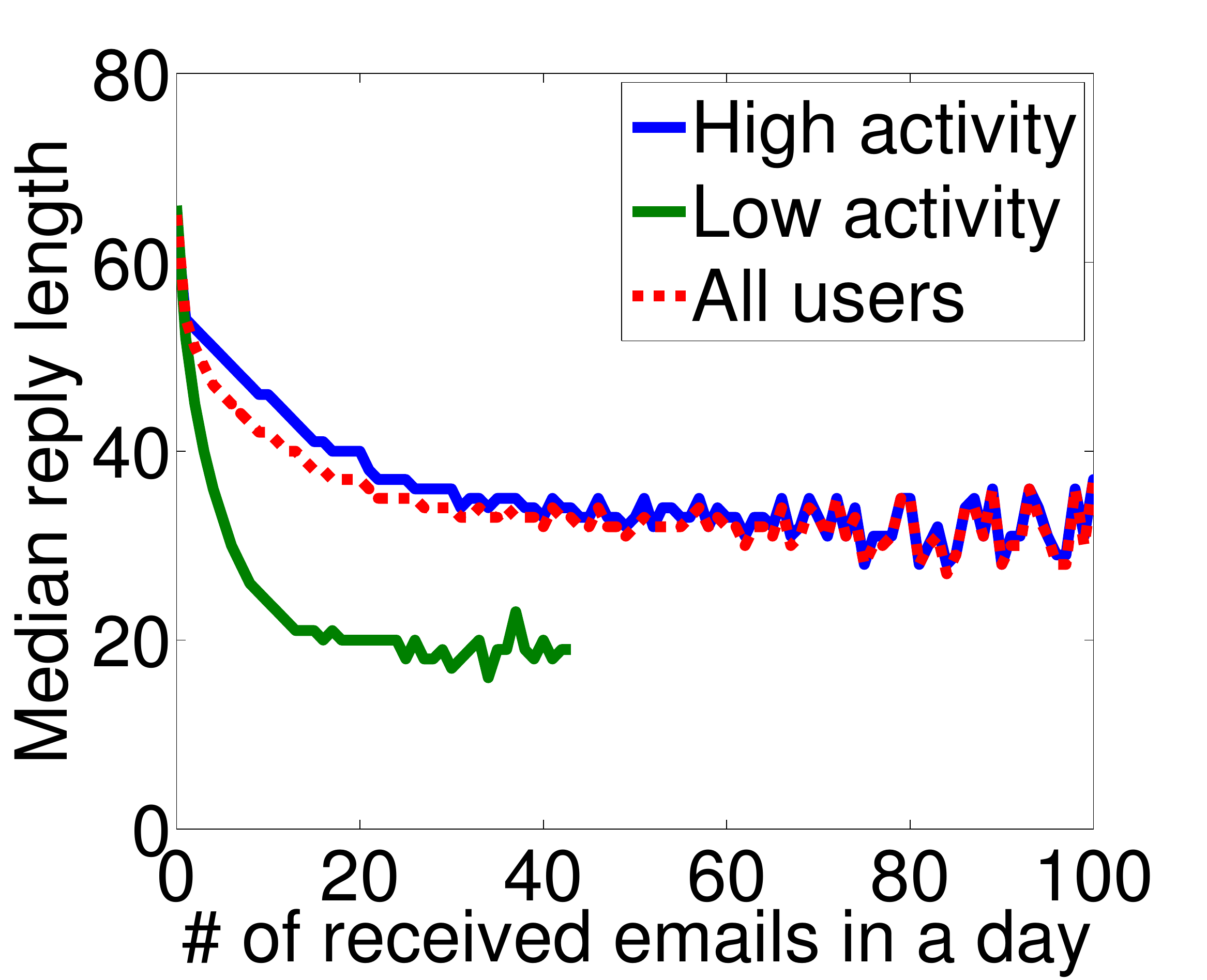}
   \label{fig:length_over_2group}
   }
  \end{tabular}
   \caption{(a) Reply time and (b) length as a function of email load. Email load is measured by the number of all emails received in a day. As email load grows, users send shorter and faster replies.}
  \label{fig:time_length_over_2group}
  \vspace{-3mm}
\end{figure}

We also investigate how email load affects reply time and length. Figure~\ref{fig:time_over_2group} shows the median reply time for messages sent on days with a given email load.
Reply time decreases rapidly as information load increases. While this may seem counterintuitive at first, it makes sense in light of an earlier study~\cite{Dabbish06}, which found that a productive strategy for reducing the perception of email overload was to frequently check the email inbox. If users check email frequently, they are more likely to respond to more recent emails, decreasing the reply time. Interestingly, low activity users reply slightly faster as email load increases.

Moreover, the length of replies also decreases as email load grows (Figure~\ref{fig:length_over_2group}). This could be partially explained as a strategy to compensate for rising information load: users send shorter messages in order to reply to more messages, and since shorter messages take less time to write, reply time decreases as well. High activity users are not affected as much by overload as low activity users. They seem to compensate and send messages that are twice as long as what the low activity users are sending.

\begin{figure}[t!]
\begin{tabular}{@{}c@{}c@{}}
\subfigure[Reply time]{
   \includegraphics[width=0.5\columnwidth]{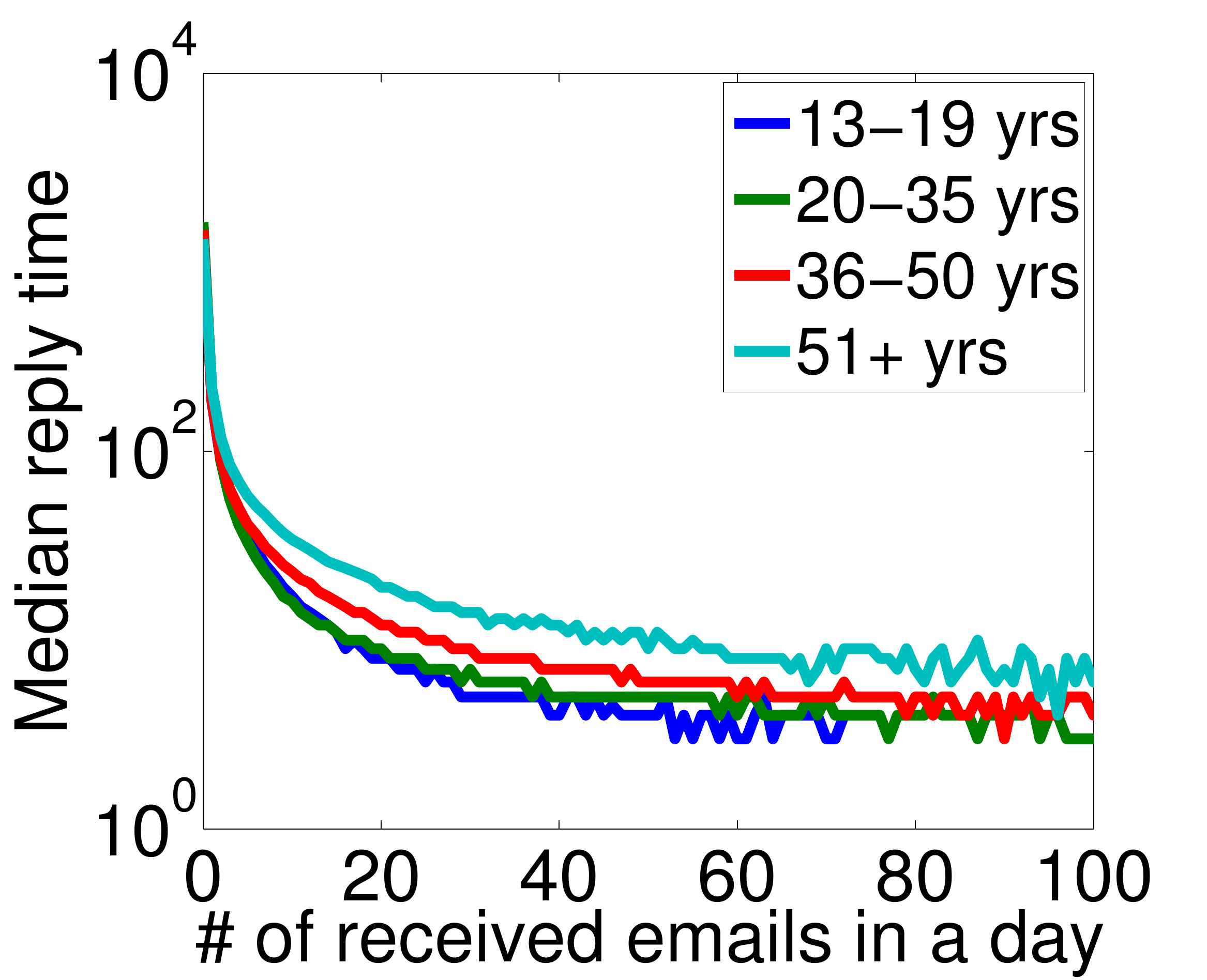}
      \label{fig:time_over_age}
   }
  &
   \subfigure[Length of reply]{
   \includegraphics[width=0.5\columnwidth]{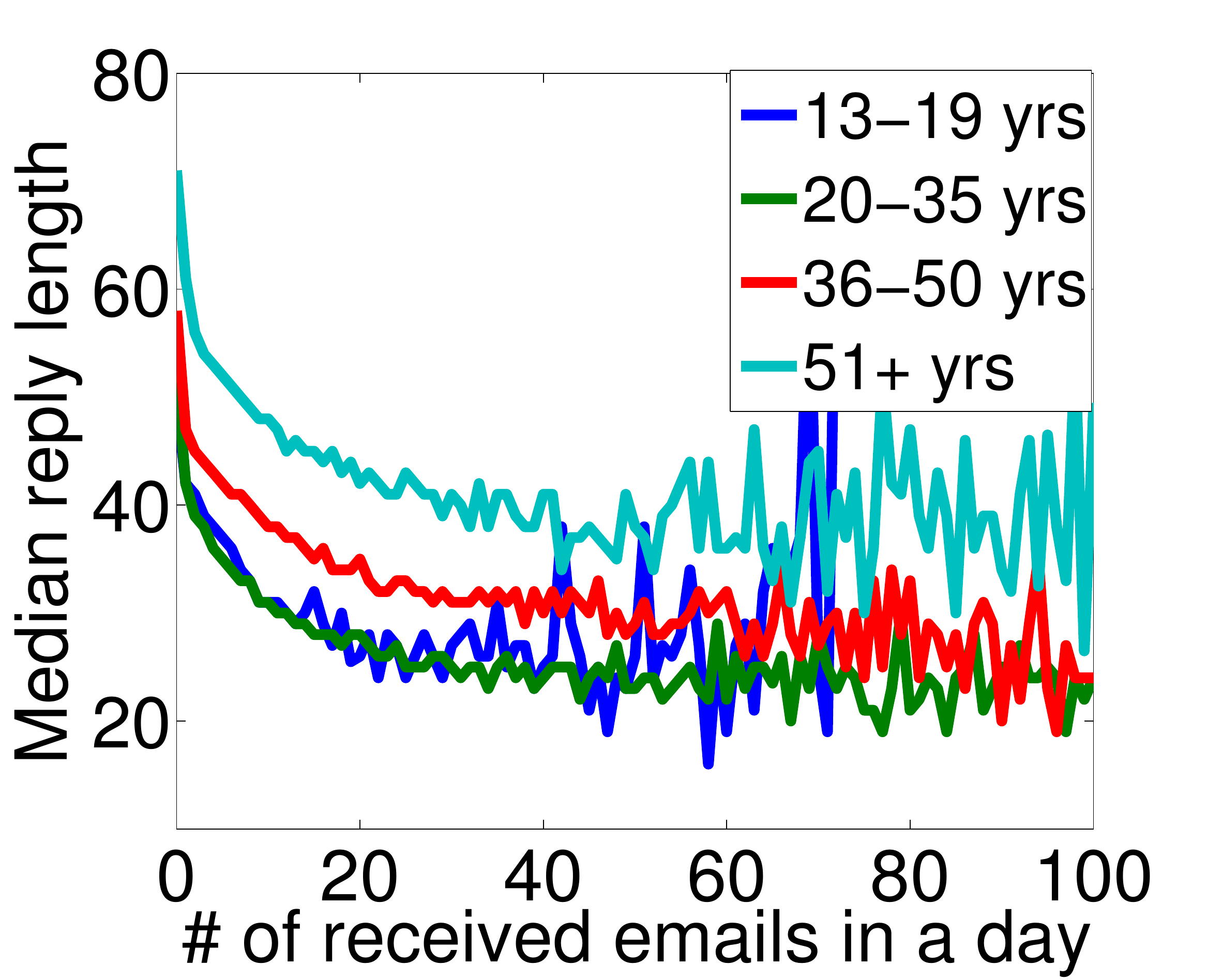}
   \label{fig:length_over_age}
   }
  \end{tabular}
   \caption{(a) Reply time and (b) length as a function of email load for users with different ages. Younger users decrease their reply time and length of reply more, to better handle the increased load.}
  \label{fig:time_length_over_age}
  \vspace{-3mm}
\end{figure}

Finally, we investigate the effect of increased load of information on reply time and length of reply of users with different ages and genders. Figure~\ref{fig:time_length_over_age} shows that younger users change their behavior more than older people, both for reply time and length of reply. When younger users become more overloaded they tend to send shorter and faster replies to cope with the increased load, on the other hand, older people are affected less with respect to reply time and length of reply, but as we have seen in the previous analyses,  older users adapt to the increased load by replying to a smaller fraction of emails. Young users and old users cope with information overload differently, younger users try to reply to as many emails as they are supposed to at the expense of spending less time for each reply and sending shorter replies, whereas older users' replies do not become significantly faster nor shorter, but they do reply to a smaller fraction of  emails.

\subsection{Synchronization of Replying Behaviors}
\noindent Do users coordinate their replies as the conversation evolves? In other words, do they become synchronized as their reply times and lengths become more similar over the course of the conversation, or do they act independently? To answer this question, we compare reply times at different stages in a thread. We do this by calculating the absolute difference in reply times in a thread, and then we divide the thread into 10 equal segments and report the mean difference within each segment. Since some users may be inherently fast, and we are interested in the change of behaviors rather than absolute value, we initially normalize the users' reply time by their median reply time. Median is used instead of the mean since reply time has a heavy-tailed distribution, and a single very long reply time would skew the mean, resulting in a very small normalized values.

\begin{figure}[t!]
\begin{tabular}{@{}c@{}c@{}}
\subfigure[Time synchronization]{
   \includegraphics[width=0.5\columnwidth]{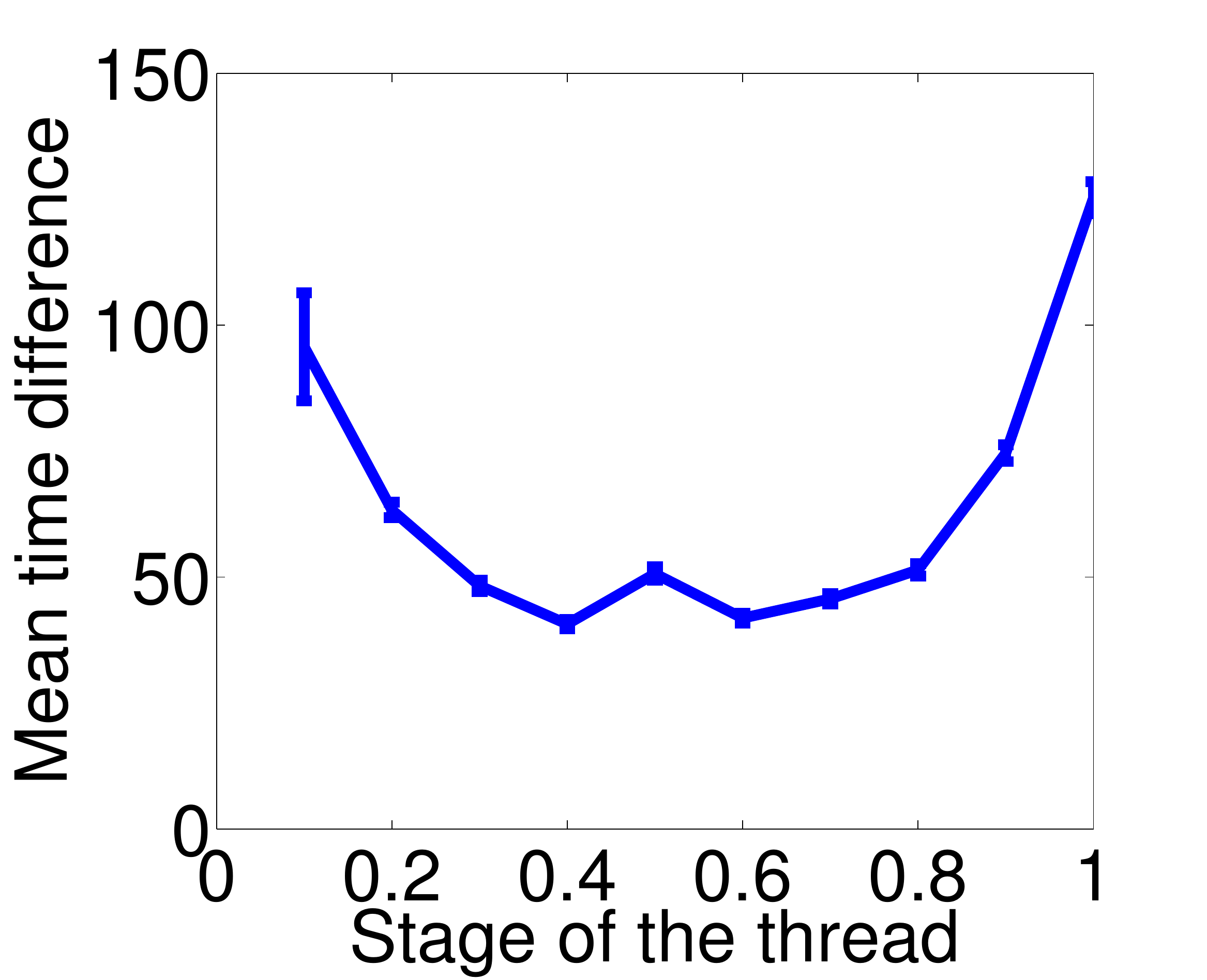}
      \label{fig:time_synch}
   }
  &
   \subfigure[Length synchronization]{
   \includegraphics[width=0.5\columnwidth]{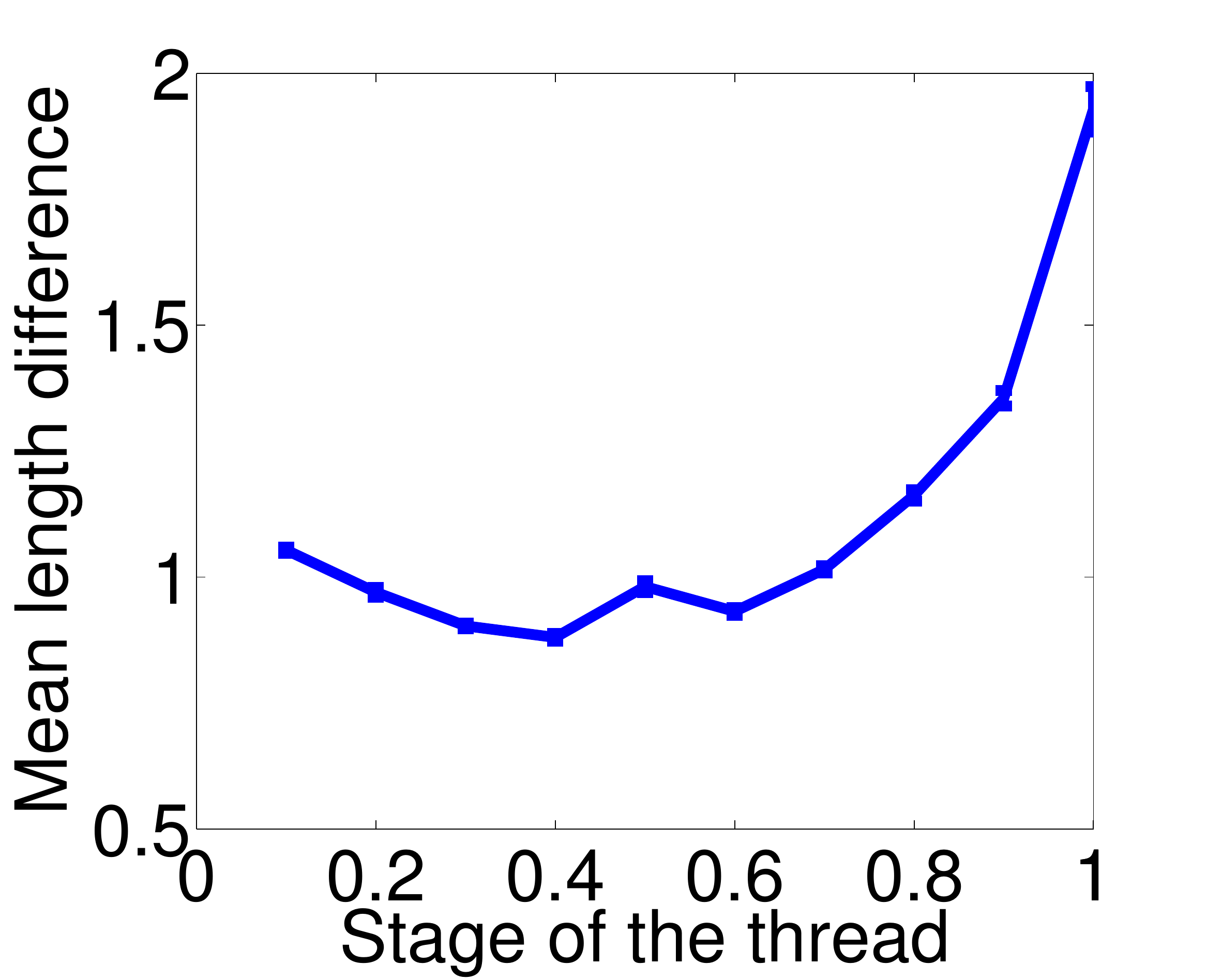}
   \label{fig:length_synch}
   }
  \end{tabular}
   \caption{Synchronization of (a) reply times and (b) length of replies over the course of the conversation with 95\% confidence interval. Users become more similar till the middle of conversation and become less similar till the end. For dyadic conversations.}
  \label{fig:time_length_synch}
  \vspace{-4mm}
\end{figure}

Figure~\ref{fig:time_synch} shows evolution of reply times in a thread, along with the 95\% confidence interval. Reply times become more similar until the middle of the conversation, and diverge afterwards: users become more synchronized until the mid-point of the conversation, then they start acting more independently. We repeat the analysis with the length of replies. Figure~\ref{fig:length_synch} shows a similar pattern: synchronization till the middle of the conversation and divergence afterwards. 
There is no significant difference in synchronization of behaviors between active and less active users.

Next, we study the content of emails in a thread, in particular, whether concepts converge over the course of a conversation. We divide each thread to 10 equal segments and calculate for each segment the cosine similarity between consecutive replies, and the average of cosine similarities at each step. 

We used a neural language model, known as paragraph2vec \cite{le2014distributed}, to represent each email as a real-valued low dimensional vector.
Introducing low dimensional embeddings of words by neural networks has significantly improved the state of the art in NLP \cite{bengio2006neural}. These take advantage of word order in documents, and state the assumption that closer words in the word sequence are statistically more dependent. 
Typically, a neural language model learns the probability distribution of next word given a fixed number of preceding words which act as the context. A recently proposed scalable Continuous Bag-of-Words (CBOW) and scalable Continuous Skip-gram (SG) model \cite{mikolov2013distributed} for learning word representations have shown promising results in capturing both syntactic and semantic word relationships in large news articles data. Their scalable open-source software is available online\footnote{code.google.com/p/word2vec}. Word representations are typically learned from large collection of documents in a sliding window-fashion by updating the central word in the window such that it is capable of accurately predicting the surrounding words in the same window. 
In a followup publication, Le and Mikolov \cite{le2014distributed} addressed an open question of how to represent a document given word vectors. They introduced a notion of a global context vector. Each document is assigned a vector of its own, which was treated as global context of all the words in that document. Word vectors and document vectors are trained simultaneously. Training is conducted in the same manner as before, in a sliding window fashion, except that the document vector is updated with every word in a sequence, as global context.

\begin{figure}[t]
\begin{center}
\includegraphics[width=0.8\columnwidth]{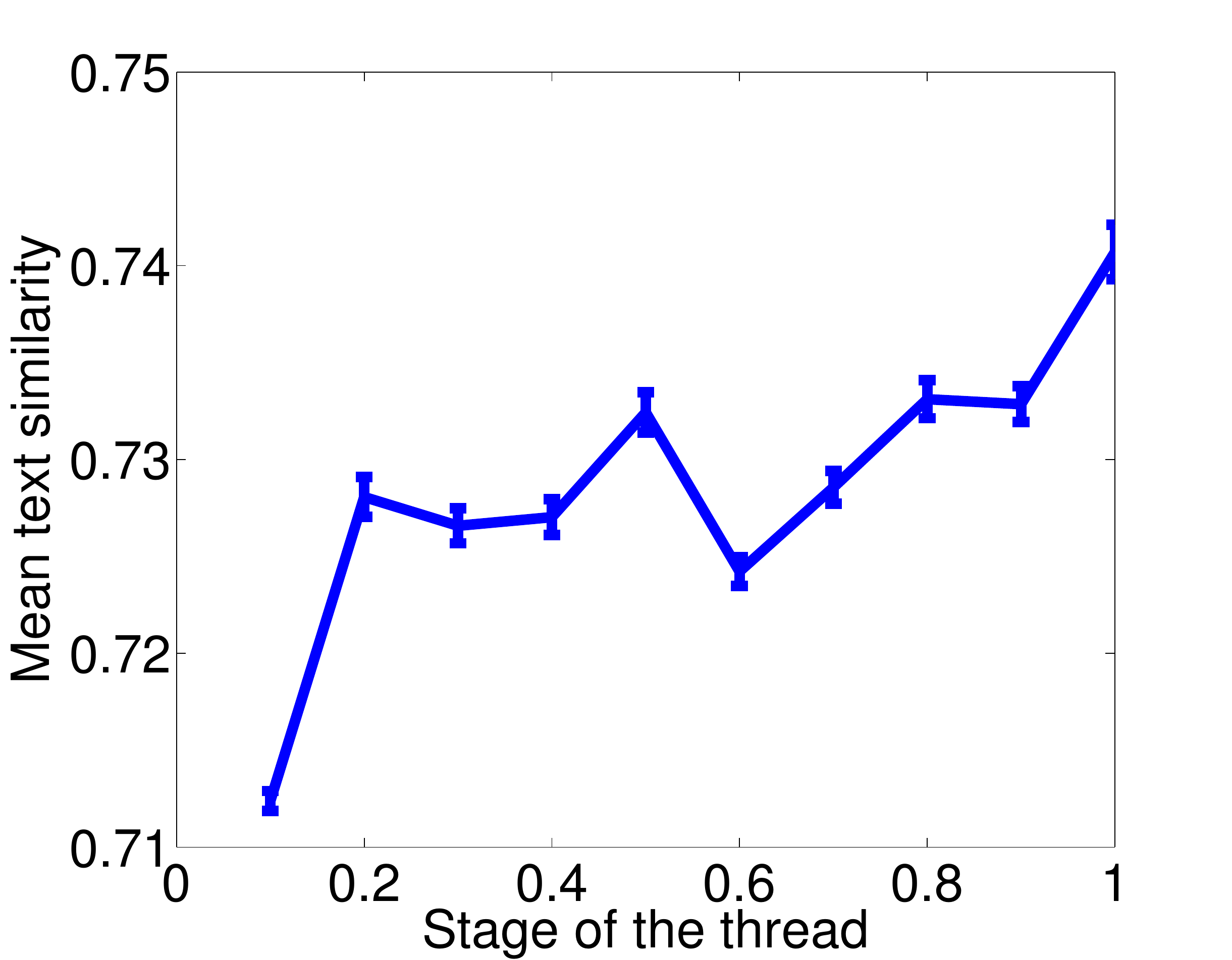}
\end{center}
\caption{Cosine similarity of body of emails represented in a lower dimension. The topics of the conversation get more similar as the conversation progresses.}
\label{fig:vector_synch}
\vspace{-3mm}
\end{figure}

We treated emails as ``documents'', and their vector representations were learned using the words in the email body. The dimensionality of the embedding space was set to $d=300$, email vectors were treated as global context, thus updated with every word in its body, while the words vectors were updated using context neighborhood of length $5$. We conducted $10$ training iterations over the entire dataset of emails and words. With the resulting vector representation, we found that the emails get more similar as the conversation progresses. Figure~\ref{fig:vector_synch} shows a considerably higher similarity of content in later stages of email threads.

Finally, we tested for linguistic style coordination. In social psychology, linguistic style coordination suggests that people mimic each other's linguistic style when they converse~\cite{danescu2012echoes}.  To quantify linguistic style, we use a set of ``markers''~\cite{danescu2012echoes}, which are specific function words that have little semantic meaning and mostly represent the style of the language. We focus on six of the well-known markers: articles, auxiliary verbs, conjunctions, personal pronouns, prepositions, and quantifiers. Conventionally, style coordination is measured by comparing the counts of markers in a statement and a response. If the count in the response is correlated with the count in the original statement, coordination occurs. But, a recent study has shown that most of the linguistic coordination can be explained by the coordination in length of the response~\cite{ShuyangGao}. In other words, a long statement probably contains many articles, and its response is likely to be long as well, containing many articles, as well. Therefore, simply comparing the counts is not enough and the actual coordination happens only when the rate of the usage of the markers is similar. To account for this, we normalize the number of markers in an email.

\begin{figure}[t!]
\begin{tabular}{@{}c@{}c@{}}
\subfigure[Articles]{
   \includegraphics[width=0.5\columnwidth]{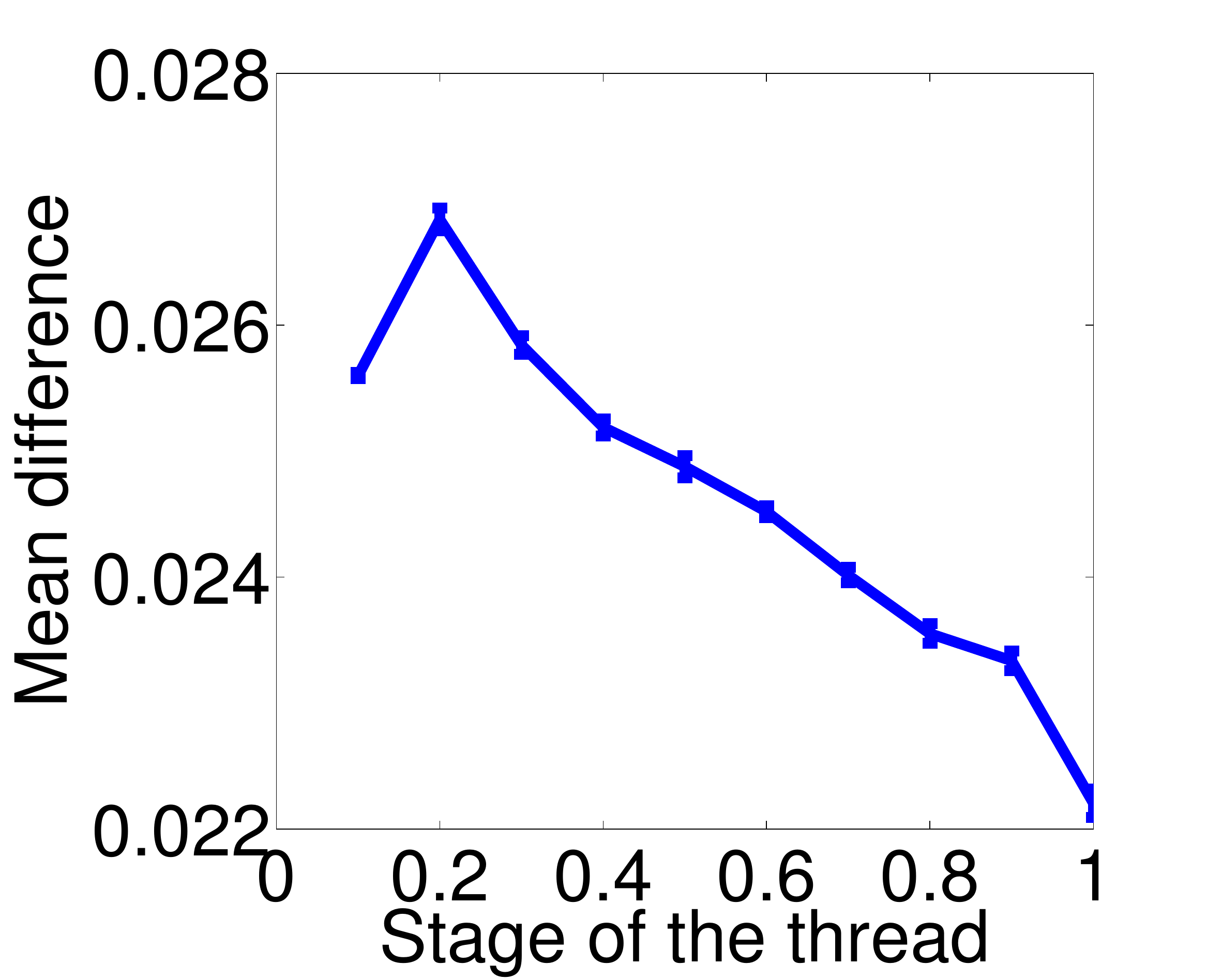}
      \label{fig:article_synch}
   }
  &
   \subfigure[Quantifiers]{
   \includegraphics[width=0.5\columnwidth]{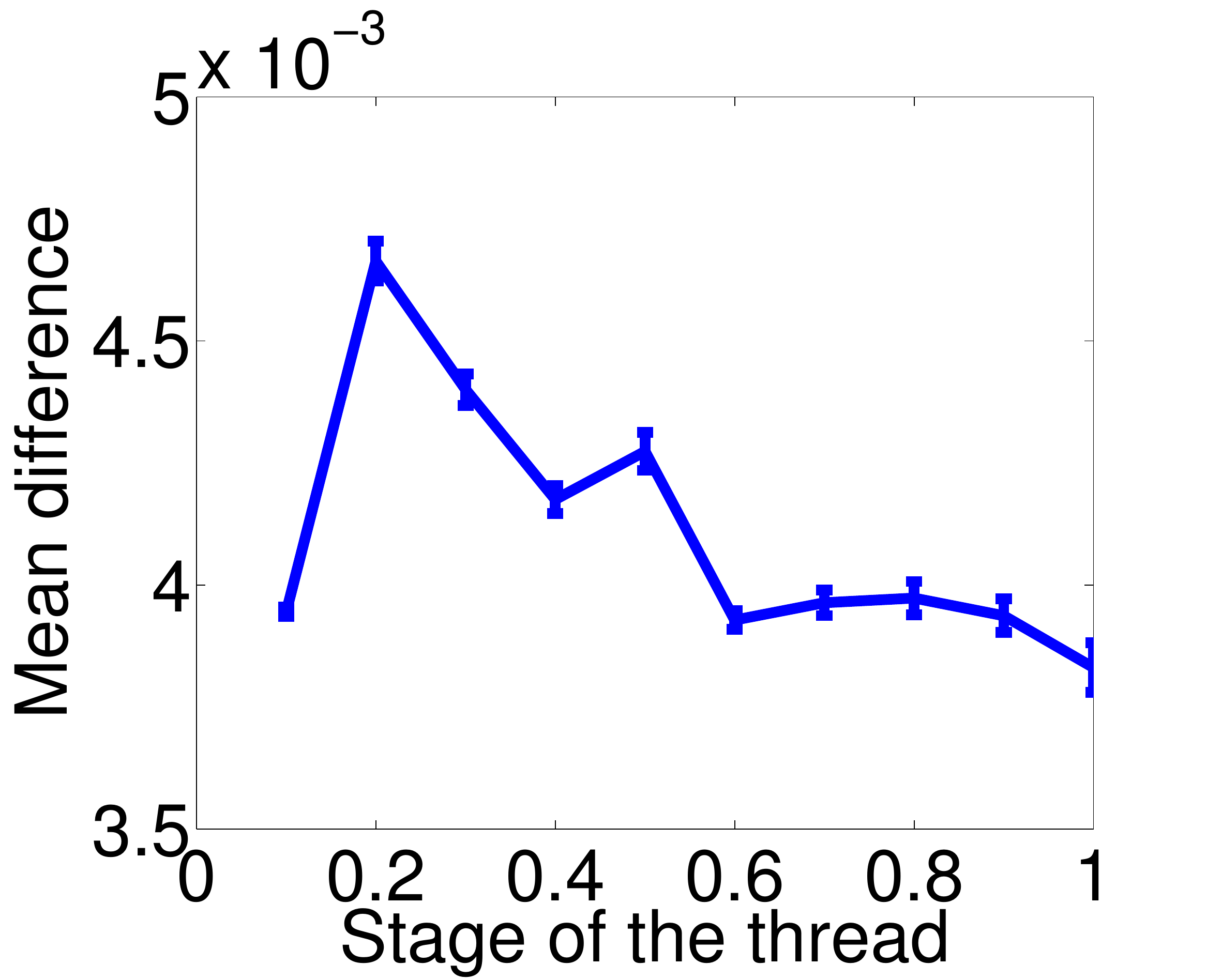}
   \label{fig:quantifier_synch}
   }
  \end{tabular}
   \caption{Synchronization in rate of usage of (a) articles and (b) quantifiers.}
  \label{fig:marker_synch}
  \vspace{-3mm}
\end{figure}

We calculate the count of the six markers, normalized by the length of email, for each email and its reply. Then, we calculate the difference in the rate of marker usage at different stages of the thread. We observe the style coordination only for two out of the six markers, articles and quantifiers (Figure~\ref{fig:marker_synch}). For articles and quantifiers the difference in the rate of marker usage decreases in a conversations, meaning that users become more similar linguistically in later stages of a thread. But this is not the case for other four markers. So, our analysis does not show a clear style coordination in language of the users, and the coordination is only happening for some of the markers.

\section{Predicting Replies}\label{section:prediction}
\noindent We test whether the features studied so far are suitable for predicting the behavior of users. In particular, we try to predict the \textit{time} that a user will take to reply to a message, the \textit{length} of the reply, and whether a message will \textit{end the thread}. More than just measuring the performance in terms of accuracy, prediction allows us to quantify the importance of features in describing emailing behavior. We use the same features for the three analysis. Features include: mean, median, and earlier reply times and reply lengths between the pair of users (20 features), age and gender of the sender and receiver (4 features), step of the thread (1 feature), statistics on number of received, sent, and replied emails for sender and receiver (18 features), statistics on number of contacts of sender and receiver (18 features), statistics on length of all emails sent and received by the sender and receiver (18 features), the time of the day and day of the week that the email was received (2 features), number of attachments (1 feature), and if the user has used a phone, or tablet earlier (1 feature). Overall, we  consider 83 features.

\subsection{Predicting Reply Time}
\noindent We start by predicting the reply time of emails within dyadic conversations. For each pair,  we use the first 75\% of the replies for training and the last 25\% for testing, so that we are not using any of the future emails for predicting the current reply time. Predicting the exact reply time is a hard problem. We simplified the problem by considering classes of replies. In practice, knowing if we are going to receive a reply shortly or with a long delay would still be very helpful and we do not necessarily need the exact reply time. We consider three balanced classes of replies: immediate replies that happen within 15 minutes (33.5\% of the data), fast replies that happen after 15 minutes, but before 164 minutes (33.1\% of the data), and slow replies that take longer than 164 minutes (33.4\% of the data). Here our baseline would be the largest class (majority vote), which contains 33.4\% of the training data. We experiment with a variety of machine learning algorithms and bagging algorithm yields the best results with Root Mean Squared Error (RMSE) of 0.420 and accuracy of 55.8\%, which is 22.4\% absolute improvement and 67.1\% relative improvement over the baseline. In case we want to just distinguish between immediate and slow replies in an application, we can eliminate our middle class and do the prediction only for the two class of immediate and slow replies. In this case, we achieve a much higher accuracy of 79.5\% compared to the baseline of 50.1\%.

We also rank the features based on their predictive power by computing the value of the $\chi^2$ statistic with respect to the class. Table~\ref{table:feature_rank_time} shows the top 5 features with the highest predictive power, and all of them come from the history of the reply times between the users. The feature with the highest predictive power is the median reply time of the replier from earlier replies. So, if we want to use only one feature for guessing the reply time of a message, the typical reply time of the replier would be the most useful feature, which makes sense. Interestingly, if we select only 7 features from all the 83 features that have high predictive power and low overlap, the accuracy would be 58.3\%, which is slightly less than the case that we consider all the features. All the 7 features represent earlier history of reply time between the pair of users.

\begin{table}[t!]
\begin {center}
{
\begin {tabular} {| c | l | r |}
\hline
{\textbf{Rank}} & {\textbf{Feature}} & \textbf{{$\chi^2$} value} \\
\hline
1 & Replier's median reply time & 6,374\\
2 & Receiver's median reply time & 4,839\\
3 & Replier's last reply time &  4,528\\
4 & Receiver's last reply time &  4,157\\
5 & Replier's 2nd to the last reply time & 3,259\\
\hline
\end{tabular}
}
\end{center}
\caption{Top 5 most predictive features for predicting reply time and their {$\chi^2$} value.}
\label{table:feature_rank_time}
\end{table}

\subsection{Predicting Reply Length}

\noindent Next, we take the same approach to predict the length of a reply that is going to be sent. We use the same set of features as the previous section and again use the first threads of emails between a pair of users for training and the rest for testing. Again, we divide our data to three balanced classes: short replies of 21 words or smaller (33.1\% of the data), medium-length replies longer than 21 words, but shorter or equal to 88 words (33.6\% of the data), and replies that are longer than 88 words (33.3\%). A naive classifier that always predicts the largest class, would have a 33.6\% accuracy, which is our baseline. Using the bagging classifier we achieve accuracy of 71.8\%, which is much higher than the prediction of reply time. Our classifier has 38.2\% absolute improvement and 113.7\% relative improvement over the baseline. Similar to time to replies, we eliminate the middle class to calculate the accuracy for distinguishing short and long replies. Our classifier can successfully assign the correct class in 89.5\% of cases, which is well above the 50.2\% baseline.

We use the $\chi^2$ statistics to rank the features based on their predictive power (Table~\ref{table:feature_rank_length}). All the top 5 features are from the earlier reply lengths of the replier. Unlike reply time, there is no feature related to the receiver's activity in the top 5 features. This suggests that the length of the reply of the other person has a weaker effect on the length of the outgoing reply, compared to the effect of the reply time of the party on the reply time of the replier. We also try the 8 features that have high predictive power and these only top 8 features are just slightly less predictive than all the features (0.2\%).

\begin{table}[t!]
\begin {center}
{
\begin {tabular} {| c | l | r |}
\hline
{\textbf{Rank}} & {\textbf{Feature}} & \textbf{{$\chi^2$} value} \\
\hline
1 & Replier's average reply length & 12,953\\
2 & Replier's last reply length &12,509\\
3 & Replier's median reply length &  11,558\\
4 & Replier's 2nd to the last reply length &  9,476\\
5 & Replier's 3rd to the last reply length & 7,595\\
\hline
\end{tabular}
}
\end{center}
\caption{Top 5 most predictive features for predicting  length of reply and their {$\chi^2$} value.}
\label{table:feature_rank_length}
\end{table}

\subsection{Predicting the End of the Thread}

\noindent Finally, we use the same approach to predict whether a reply is the last reply in a thread or not. The baseline for this prediction problem is 50.6\% and bagging classification yields accuracy of 65.9\%, i.e. 15.3\% absolute improvement and 30.2\% relative improvement over the baseline. Table~\ref{table:feature_rank_last} shows the top 5 predictive features for predicting the last email in a thread and interestingly all the top features are related to the load of information in terms of number of words on the replier or receiver.

\begin{table}[t!]
\begin {center}
{
\begin {tabular} {| c | l | r |}
\hline
{Rank} & {Feature} & {$\chi^2$} value \\
\hline
1 & Receiver's avg \# of words received/day & 2,160\\
2 & Replier's avg \# of words received/day & 1,981\\
3 & Receiver's median \# of words received/day &  1,935\\
4 & Receiver's avg \# of words sent/day &  1,884\\
5 & Replier's median \# of words received/day& 1,872\\
\hline
\end{tabular}
}
\end{center}
\caption{Top 5 most predictive features for predicting last email in a thread and their {$\chi^2$} value.}
\label{table:feature_rank_last}
\end{table}

Table~\ref{table:prediction_results} summarizes our results for the three prediction problems. Besides the majority vote baseline, we also considered last reply and most used reply time and length as other baselines. These baselines perform better than the majority vote, but our classifier outperforms all three baselines: Relative improvement is 17.1\% for last reply time and 5.3\% for the last reply length. For most used baseline the relative improvement is 30.4\% for reply time and 58.9\% for reply length.

\begin{table*}[t!]
\begin {center}
{
\begin {tabular} {| l | c | c | c | c | c | c | c | c |}
\hline
{Prediction} &  {Majority vote} & {Last reply} & {Most used} & {Our classifier} &  \specialcell{Absolute \\ improvement} & \specialcell{Relative \\ improvement} & AUC & RMSE\\
\hline
Reply time & 33.4\% &50.2\% & 45.1\% & 58.8\% & 22.4\% & 67.1\% & 0.715 & 0.420\\
Reply length & 33.6\% & 68.2\% & 45.2\% & 71.8\% & 38.2\% & 113.7\%& 0.865 & 0.361\\
Last email & 50.6\% & -- & -- & 65.9\% & 15.3\% & 30.2\%& 0.761 & 0.454\\
\hline
\end{tabular}
}
\end{center}
\caption{Summary of the prediction results. Accuracy: percentage of correctly classified samples. AUC: Weighted average of Area Under the Curve for classes. RMSE: Root Mean Square Error. The improvements are reported over the majority vote baseline.}
\label{table:prediction_results}
\end{table*}

\section{Related Work}\label{section:related}

\noindent \textbf{Online conversations} were studied extensively in the context of social media, as they are important drivers of user \textit{engagement} in online communities~\cite{harper07talk}. Research tried to identify common predictive models of the main traits of human communication. Previous work on Twitter  investigated the \textit{conventions} used to initiate and track conversations~\cite{boyd10tweet}. The evolution of conversational \textit{norms} in time have been recently studied on Flickr and aNobii~\cite{aiello14reading}. Similar to our study, previous work tried to predict the length of a discussion thread in Facebook using time and content features~\cite{backstrom2013}. Models to reproduce some statistical properties of threads (e.g., size of thread, number of participants) were tested successfully in Twitter and Yahoo Groups~\cite{kumar10dynamics}. Multimodal features of discussion threads can also predict its perceived interestingness~\cite{dechoudhury09conversations}. 
Divergence of topics in Twitter group conversations was recently investigated~\cite{purohit14understanding}. 
The propensity to engage in conversation has been investigated under the light of the user personality traits such as openness to new experiences or emotional stability~\cite{correa10who,celli12role}. The emotions conveyed in online conversations were also studied~\cite{java07why,kim12feel,budak13participation}. We believe ours is the first large scale analysis of email conversations and attempt to predict some of their main structural properties. Also, we do not focus on properties of the agents involved in conversations except for some demographic features like age and gender.


\vspace{5pt} \noindent \textbf{Emailing behavior} was studied predominantly on small-scale data and often using qualitative methods. The attitude of email users towards work email was investigated through organizational surveys~\cite{dabbish05understanding}, finding that the social nature of the message is a stronger motivation to reply than the ``importance'' of the message. Also, survey respondents tended to reply to about only a third of the messages in their inbox. User studies and targeted interviews about rhythms in email usage, including intervals of replying, have uncovered the role of user expectation in relation with the replying behavior~\cite{tyler03response}. In particular, when a user perceives that the response has been delayed too much, the resulting \textit{breakdown perception} triggers a follow-up action in the thread.

Quantitative studies targeted to specific application-oriented tasks such as classification of emails into folders~\cite{shetty04enron,klimt04enron} were conducted mainly on the open Enron email data~\cite{klimt04introducing}, which is one of the few complete temporal data on email communication whose structure has been studied extensively~\cite{diesner05communication}. However, Enron's email communication patterns were very specific as they were limited to the context of the company and by its evolution and dramatic fall. This peculiarity held back researchers from drawing conclusions on general patterns of email use. Other less known, small-scale email datasets, such as the email corpora from OSS projects~\cite{bird06mining} can provide interesting traces on user interaction, but they do not allow any generalization of the findings. Email communication networks have been also used to investigate the propagation of computer viruses~\cite{newman02emailnetworks} and word-of-mouth advertising~\cite{phelps04viral}, or to solve specific tasks such as expert finding within organizations~\cite{campbell03expertise}.

Information overload in email was studied since the 90's. In contrast to Whittaker and Sidner~\cite{Whittaker96}, who defined  \textit{email overload} as the use of email as a tool for task management, archival, and communication, we follow the definition of overload as not being able to keep up with the volume of incoming email~\cite{Dabbish06,fisher06revisiting}. Unlike previous qualitative studies that focused on the perception of overload~\cite{Dabbish06}, we focus on quantitative measures of overload and its observed effects on users' behavior.


\section{Conclusion}\label{section:conclusion}

\noindent While email accounts for a considerable portion of interpersonal communication, emailing behavior is not well understood. We carried out a large-scale study of email replying behavior of more than 2M users. We studied how a variety of factors affects reply time and length. We found that users reply faster to emails received during weekdays and working hours, and that replies tend to become shorter later in the day and on weekends. In regard to demographics, younger users generally send faster and shorter replies, and men send slightly faster and shorter replies than women. Among other factors, replies from mobile devices were faster and shorter than from desktops, and emails without attachments typically got faster replies.

We investigated the effect of email overload on the replying behavior. We found that users increased their activity as they received more emails, but not enough to compensate for the higher load. This means that as users became more overloaded, they replied to a smaller fraction of incoming emails and with shorter replies. However, their responsiveness remained intact and may even be faster. Demographic factors affected information overload, too. Older users generally replied to a smaller fraction of incoming emails, but their reply time and length were not impacted by overload as much as younger users. In contrast, younger users replied faster, but with shorter replies and to a higher fraction of emails.

We also studied synchronization of replying behavior within a thread, and found that users tended to become more similar, both in reply time and length, until the middle of a thread. After that, their behavior became less similar. We also tested for coordination of linguistic styles using a variety of markers. Results were inconclusive: some markers suggested linguistic style coordination, but this was not the case for all.

Finally, to measure the predictive power of considered factors, we built classifiers to predict the time and length of replies, and whether an email was the last one in a thread. We obtained accuracy of 58.8\%, 71.8\%, and 65.9\% for these tasks, which represented a relative improvement of 67.1\%, 113.7\%, and 30.2\% respectively over the baselines. Our classifiers could be used to improve email client applications, to better classify and rank emails. 

\subsection*{Acknowledgements}
This work was supported in part by AFOSR (contract FA9550-10-1-0569), by DARPA (contract W911NF-12-1-0034), and by the NSF under grant SMA-1360058.
\newpage

\balance
\bibliographystyle{abbrv}
{
\small
\bibliography{email_norm}
}

\end{document}